\pdfoutput=1

\documentclass{jfm}
\usepackage{color,soul}
\usepackage{graphicx}
\usepackage{amsmath}
\usepackage{booktabs}
\usepackage{amssymb}

\renewcommand{\emph} [1] {{#1}}
\input{tcilatex}

\begin{document}

\title[Strong swirl approximation and atmospheric vortices]{\emph{Strong swirl approximation \\ and \\ intensive vortices in the atmosphere}}

\author[A. Y. Klimenko]%
{A. Y. Klimenko%
  \thanks{Email address for correspondence: klimenko@mech.uq.edu.au}}

\affiliation{The University of Queensland, SoMME 
 Qld, 4072, Australia}

\pubyear{2012}
\volume{000}
\pagerange{000--000}
\date{J. Fluid Mech. January, 2014, vol. 738, pp. 268--298}

\maketitle

\bigskip

\bigskip
\bigskip

\begin{abstract}

This work investigates intensive vortices, which are characterised by the existence of a converging radial flow that significantly intensifies the flow rotation. Evolution and amplification of the vorticity present in the flow play important roles in
the formation of the vortex. When rotation in the flow becomes sufficiently strong --- and this implies validity of the strong swirl approximation developed by Einstein and Li (1951), Lewellen (1962), Turner (1966) and Lundgren (1985) --- the analysis of Klimenko (2001a-c) and of the present work determine that further amplification of vorticity is moderated by interactions of vorticity and velocity. This imposes physical constraints on the flow resulting in the so-called compensating regime, where the radial distribution of the axial vorticity is characterised by the 4/3 and 3/2 power laws. This asymptotic treatment of a strong swirl is based on vorticity equations and involves higher order terms. \emph{ This treatment incorporates multiscale analysis indicating downstream relaxation of the flow to the compensating regime. The present work also investigates and takes into account viscous and transient effects.} One of the main points of this work is the applicability of the
power laws of the compensating regime to intermediate regions in large atmospheric vortices, such as tropical cyclones and tornadoes.

\end{abstract}

\newpage





\section{Introduction}

Vortices with intense rotation occur in nature at very different scales,
with the bathtub vortex representing one of the smallest and atmospheric
vortices -- tornadoes, mesocyclones and cyclones -- representing vortices of
much greater scales. Fujita (1981), in his classical work on vortices in
planetary atmospheres, introduced a unified treatment of the vortical motion
of different scales starting from a lab vortex (that is referred to here as
a bathtub vortex) and finishing with the largest known vortex of that time
-- the Jovian Great Red Spot, whose size exceeds the Earth diameter. (\emph{%
While the Great Red Spot is not one of the intensive cyclonic vortices,
which are of interest in the present work, the polar vortices on Saturn,
which have been recently filmed by the Cassini spacecraft and are also
comparable to the size of our planet, do bear some resemblance to
terrestrial hurricanes --- see Dyudina et al., 2008).} The vortices were
classified according to their scales, and vortical motions of this kind are
viewed by Fujita as a truly universal feature of nature. The similarity
between vortices of different scales is determined by the conservation of
angular momentum principle, that is rotation must intensify as fluid moves
towards the centre of the flow. In the present work, the vortices of this
kind are referred to as \textit{intensive}. In spite of this similarity,
intensive vortices of different scales are, generally, different phenomena
characterised not only by different scales but also by different levels of
buoyancy, turbulence and axial symmetry present in the flow. There are
obvious geometrical differences between these vortices: tornadoes are tall,
column-like vortices while tropical cyclones are flat disks covering large
areas. Thus, although it is unlikely that any common approach can fully
characterise the whole structure of the vortices, this does not eliminate
the possibility of finding common explanations for certain features of the
vortices even if they represent different phenomena. Similarities and
differences of lab vortices and atmospheric vortices have been repeatedly
discussed in other publications (Turner and Lilly, 1963; Church and Snow,
1993; van Heijst and Clercx, 2009).

Observations of intensive vortices in a bathtub indicate that a) the
vortices seem to be axisymmeric, b) the Reynolds number in the flow is,
typically, very high (although the main part of the flow tends to remain
laminar), c) the evolution of the flow is quite slow compared to its intense
rotation and d) density practically remains constant over a large range of
radii. After a short initial period of unsteadiness the vortex becomes
quasi-steady. Although any results obtained on the basis of these
assumptions can not be expected to reproduce all characteristics of complex
atmospheric vortices, one can hope that some similarities can be found in a
selected region of the flows. \emph{In the region of interest, which is
intermediate between the inner (core) and outer scales of the vortex, the
axial vorticity is greatly intensified as fluid flows towards the centre.
This region is called here the \textit{intensification region}. }In
application to atmospheric vortices, using these assumptions within selected
regions was repeatedly considered in the literature (Gray, 1973; Lewellen,
1993).

The present analysis of vortical flows in a bathtub follows the strong swirl
approximation introduced by Einstein and Li (1951), Lewellen (1962), Turner
(1966) and Lundgren (1985). Another family of self-similar vortical
solutions has been obtained by Long (1961) and generalised by
Fernandez-Feria et al. (1995). Although these solutions are undoubtedly
interesting, they, as remarked by Turner (1966), are different from the
vortices considered here.

\emph{Intense swirls have been investigated theoretically and experimentally
for confined vortices by Escudier et al. (1982) and for helical vortices by
Alekseenko et al. (1999). Escudier et al. (1982) emphasise importance of
axial vorticity in the flow surrounding the core of the vortex, while the
helical approximation of Alekseenko et al. (1999) has essentially non-zero
tangential components of the vorticity. These features of the vortical flows
are congenial with the present analysis. There is, however, an essential
difference: the vortical approximations of Escudier et al. (1982) and
Alekseenko et al. (1999) were developed primarily for confined vortical
flows, while the analysis of this work is directed towards flows with
asymptotically large ratios of the outer and inner scales, such as
unconfined flows occurring in large atmospheric vortices. }

\subsection{\emph{Vortices in the atmosphere}}

Tropical cyclones (hurricanes and typhoons) have been analysed in a large
number of publications. Only some reviews of this topic --- Gray (1973),
Lighthill (1998), Emanuel (2003) and Chan (2005) --- are mentioned here.
Tropical cyclones are formed over warm oceans and act like a heat engine
converting internal energy into kinetic energy of the hurricane winds.
Compared to tornadoes, the flow pattern in tropical cyclones is more regular
and cyclones can persist for many days. The outer diameter of a strong
cyclone can reach 500 --- 1000km (Chan, 2005) which, according to Holland
(1995), is associated with the tropical Rossby length or with characteristic
synoptic scales. The influence of cyclone winds can be detected at distances
reaching 1000km from its centre (Emanuel, 2003).

The structure of tornadoes has been repeatedly reviewed in publications
(Fujita, 1981; Lewellen, 1993; Vanyo, 1993; Davies-Jones et al., 2001). In
some publications (see Lewellen, 1993), tornadoes are discussed in terms of
axisymmetric flows with viscous effects being enhanced by the presence of
atmospheric turbulence. This approach is most applicable to the core region
of tornadic flows which is associated with significant influence of the
turbulent viscosity. An alternative treatment of tornadoes (see Davies-Jones
et al., 2001) is based on inviscid analysis that takes into account
non-axisymmetric effects and seems to be most relevant to the processes
originated at larger scales. The largest scales of a supercell tornado
correspond to the core of the parent mesocyclone and are associated with
buoyancy and latent energy of atmospheric storms (Klemp, 1987). These two
major theoretical approaches are not necessarily contradictory as they can
be applicable to different regions of the tornadic motion. Asymptotic
treatment of this interpretation implies the existence of an overlap region,
which is termed here the intensification region and is of interest in the
present work.

\emph{Firewhirls are vortices driven by buoyant forces induced by the heat
released in large fires (Williams, 1982). Rotation in firewhirls is intense
and this tends to further stimulate the fires that become very difficult to
extinguish. In some firewhirls, the vortex is so strong that even the
buoyant forces start to play a secondary role: inclined firewhirls have been
repeatedly observed in nature and in experiments (Chuah et al., 2011). }

\subsection{\emph{Outline of the present work}}

Section 2 introduces the major equations and dimensionless groups governing
intensive vortical flows qualitatively similar to bathtub vortex. The main
feature of the present approach is its emphasis on the evolution of
vorticity, resulting in a bias towards using vorticity (Helmholtz) rather
than velocity (Navier-Stokes) equations. If rotation in the vortical flow
remains relatively weak, then its complete description is easy: the flow on
the planes passing through the axis must be potential (or close to
potential). \emph{This case is referred to as a vortex with a potential
axial-radial flow image and should be distinguished from the conventional
two-dimensional flow called potential vortex. The Burgers (1940) vortex is a
good example of a vortex with a potential axial-radial image.} We, of
course, are interested in the case of relatively fast rotation in the flow,
which is much more complicated and relevant to the realistic vortices
observed in a bathtub and in the atmosphere. Analysis of axisymmetric flows
with strong vorticity is performed in Section 3, where generic bathtub-like
vortices are considered and special attention is paid to the intensification
region. Similar to Lundgren (1985), the vortices are treated here as
axisymmetric and incompressible flows with intense rotation and low
viscosity (viscous effects nevertheless still can play a significant role in
some regions), although the present analysis involves higher-order
expansions characterising strong velocity/vorticity interactions.

Among atmospheric vortices, tropical cyclones (hurricanes and typhoons) and
strong tornadoes are characterised by their most distinct signatures. The
ability of the suggested theory to adequately represent certain features of
cyclones, tornadoes and other vortices is investigated in Section 4. In this
section, several examples of the radial distribution of axial vorticity
reported in the literature for atmospheric vortices are shown demonstrating
a reasonably good agreement with the theory. The results recently obtained
by Klimenko and Williams (2013) for firewhirls are discussed in the context
of the presented approach. The Appendix presents mathematical details of the
asymptotic analysis of viscous effects in the core of the vortex and of the
unsteady vorticity evolution.

\emph{The present theory generalises the previous analysis of Klimenko
(2001a-c, 2007). This work introduces viscous terms into the analysis and
demonstrates that the singularity of the inviscid solution disappears within
the viscous core; performs the multi-scale analysis of and gives a physical
interpretation for the vortical relaxation mechanism that balances the
values of the exponents; and finally investigates evolution of the strong
vortices and examines applicability of the developed theory to intensive
vortices in the atmosphere. Most impotently, the present theory is shown to
be in excellent agreement with the most comprehensive investigation of
vorticity distribution in tropical storms by Mallen et al. (2005). }

\section{Axisymmetric vortical flows}

The present consideration begins with a generic vortical flow which, as
discussed in the Introduction, is assumed to be axisymmetric and
incompressible. The flow is characterised by intense rotation resembling
that of a bathtub vortex, although fluid flows downwards in bathtub vortices
and upwards in atmospheric vortices. A conventional cylindrical system of
coordinates $r$, $z$, $\theta $ is used here with the positive direction of
the $z$-axis selected along the direction of the axial flow. \emph{Since
axial vorticity is deemed to be present in the flow surrounding the vortex,
the centripetal motion amplifies the axial vorticity by axial stretch and
intensifies rotation due to conservation of angular momentum. The vortex
intensifies and evolves in time. Since this work is not interested in prompt
or sudden changes, the unsteady effects are considered only when they are
intrinsic to the flow. The vortex is thus seen as quasi-steady and
preserving its symmetry. }

\subsection{Governing equations}

The following form of the incompressible Navier-Stokes equation 
\begin{equation}
\frac{\partial \mathbf{v}}{\partial t}+\mathbf{\nabla }B=\mathbf{v}\times 
\mathbf{\omega }+\nu \nabla ^{2}\mathbf{v,\ \mathbf{\nabla }\cdot v}=0,\ 
\mathbf{\ }B\equiv \frac{v^{2}}{2}+\frac{p}{\rho }+gz,\ \mathbf{\ \omega }%
\equiv \mathbf{\mathbf{\nabla }\times v}  \label{11NS}
\end{equation}
is most convenient for the analysis. Here, $\mathbf{v}$ is velocity, $%
\mathbf{\omega }$ is vorticity, $B$ is the Bernoulli integral and the sign
of $g$ takes into account direction of the gravity with respect to the
direction of the vertical axis $z$. In a laminar flow $\nu $ denotes
molecular viscosity but, if turbulence is present in the flow, $\nu $ should
be treated as the effective turbulent viscosity.\ The axisymmetric ($%
\partial /\partial \theta =0$) form of these equations is given by Batchelor
(1967): 
\begin{equation}
v_{r}\omega _{\theta }-v_{\theta }\omega _{r}=\frac{\partial v_{z}}{\partial
t}+\frac{\partial B}{\partial z}-\nu \left( \frac{1}{r}\frac{\partial }{%
\partial r}\left( r\frac{\partial v_{z}}{\partial r}\right) +\frac{\partial
^{2}v_{z}}{\partial z^{2}}\right) ,  \label{11NSz}
\end{equation}
\begin{equation}
v_{\theta }\omega _{z}-v_{z}\omega _{\theta }=\frac{\partial v_{r}}{\partial
t}+\frac{\partial B}{\partial r}-\nu \left( \frac{\partial }{\partial r}%
\left( \frac{1}{r}\frac{\partial v_{r}r}{\partial r}\right) +\frac{\partial
^{2}v_{r}}{\partial z^{2}}\right) ,  \label{11NSr}
\end{equation}
\begin{equation}
v_{z}\omega _{r}-v_{z}\omega _{z}=\frac{\partial v_{\theta }}{\partial t}%
-\nu \left( \frac{\partial }{\partial r}\left( \frac{1}{r}\frac{\partial
v_{\theta }r}{\partial r}\right) +\frac{\partial ^{2}v_{\theta }}{\partial
z^{2}}\right) ,  \label{11NStet}
\end{equation}
\begin{equation}
\frac{\partial v_{r}r}{\partial r}+\frac{\partial v_{z}r}{\partial z}=0,\ \
\omega _{\theta }=\frac{\partial v_{r}}{\partial z}-\frac{\partial v_{z}}{%
\partial r},\quad \omega _{r}=-\frac{\partial v_{\theta }}{\partial z},\quad
\omega _{z}=\frac{1}{r}\frac{\partial v_{\theta }r}{\partial r},
\label{11NSdiv}
\end{equation}
where $r$, $z$ and $\theta $ are the conventional cylindrical coordinates
and, as subscript indices, denote the corresponding components of the
vectors. With the use of substantial derivative $\mathrm{d}/\mathrm{d}t,$
the stream function $\psi $ and the circulation $2\pi \gamma $%
\begin{equation}
\frac{\mathrm{d}}{\mathrm{d}t}\equiv \frac{\partial }{\partial t}+v_{z}\frac{%
\partial }{\partial z}+v_{r}\frac{\partial }{\partial r},\;v_{z}=\frac{1}{r}%
\frac{\partial \psi }{\partial r},\;v_{r}=-\frac{1}{r}\frac{\partial \psi }{%
\partial z},\ \ \gamma \equiv v_{\theta }r,  \label{11dt}
\end{equation}
the system of governing equations can be written in the form 
\begin{equation}
\frac{\partial ^{2}\psi }{\partial z^{2}}+r\frac{\partial }{\partial r}%
\left( \frac{1}{r}\frac{\partial \psi }{\partial r}\right) =-r\omega
_{\theta },  \label{11psi}
\end{equation}
\begin{equation}
\frac{\mathrm{d}\omega _{\theta }/r}{\mathrm{d}t}-\nu \left( \frac{1}{r}%
\frac{\partial }{\partial r}\left( \frac{1}{r}\frac{\partial \omega _{\theta
}r}{\partial r}\right) +\frac{\partial ^{2}\omega _{\theta }/r}{\partial
z^{2}}\right) =-2\frac{\gamma \omega _{r}}{r^{3}},  \label{11wt}
\end{equation}
\begin{equation}
\frac{\mathrm{d}\gamma }{\mathrm{d}t}=\nu \left( r\frac{\partial }{\partial r%
}\left( \frac{1}{r}\frac{\partial \gamma }{\partial r}\right) +\frac{%
\partial ^{2}\gamma }{\partial z^{2}}\right) \;.  \label{11g}
\end{equation}
\emph{In the rest of the paper, the value $\gamma =v_{\theta }r,$ which is
different from the conventional definition of this quantity by the factor of 
$2\pi $, is referred to as circulation.} Here, the equation for $\omega
_{\theta }$ is obtained by differentiating (\ref{11NSz}) with respect to $r,$
differentiating (\ref{11NSr}) with respect to $z$ and subtracting the
results while taking into account the following equations 
\begin{equation*}
\ \mathbf{\omega }\cdot \mathbf{\nabla }\gamma =\omega _{z}\frac{\partial
\gamma }{\partial z}+\omega _{r}\frac{\partial \gamma }{\partial r}=0,\ \ 
\mathbf{\omega }\cdot \mathbf{\nabla (}\gamma r^{-2})=\gamma \mathbf{\omega }%
\cdot \mathbf{\nabla }r^{-2}=-2\frac{\gamma \omega _{r}}{r^{3}}=\frac{1}{%
r^{4}}\frac{\partial \gamma ^{2}}{\partial z},
\end{equation*}
\begin{equation}
\omega _{r}=-\frac{1}{r}\frac{\partial \gamma }{\partial z};\quad \omega
_{z}=\frac{1}{r}\frac{\partial \gamma }{\partial r},\;\ \ \frac{\partial
\omega _{r}r}{\partial r}+\frac{\partial \omega _{z}r}{\partial z}=0\;.
\label{11ww}
\end{equation}
The equations for the vorticity components $\omega _{z}$ and $\omega _{r}$
can be easily obtained by applying the operators $\partial /\partial r$ and $%
\partial /\partial z$ to equation (\ref{11g}) 
\begin{eqnarray}
\frac{\partial \omega _{z}r}{\partial t}+\frac{\partial \left( v_{r}\omega
_{z}-v_{z}\omega _{r}\right) r}{\partial r} &=&  \notag \\
r\frac{\mathrm{d}\omega _{z}}{\mathrm{d}t}-r\omega _{r}\frac{\partial v_{z}}{%
\partial r}-r\omega _{z}\frac{\partial v_{z}}{\partial z} &=&\nu \left( 
\frac{\partial }{\partial r}\left( r\frac{\partial \omega _{z}}{\partial r}%
\right) +\frac{\partial ^{2}\omega _{z}r}{\partial z^{2}}\right) ,
\label{11wz}
\end{eqnarray}
\begin{eqnarray}
\frac{\partial \omega _{r}r}{\partial t}+\frac{\partial \left( v_{z}\omega
_{r}-v_{r}\omega _{z}\right) r}{\partial z} &=&  \notag \\
r\frac{\mathrm{d}\omega _{r}}{\mathrm{d}t}-r\omega _{r}\frac{\partial v_{r}}{%
\partial r}-r\omega _{z}\frac{\partial v_{r}}{\partial z} &=&\nu \left( r%
\frac{\partial }{\partial r}\left( \frac{1}{r}\frac{\partial \omega _{r}r}{%
\partial r}\right) +\frac{\partial ^{2}\omega _{r}r}{\partial z^{2}}\right) .
\label{11wr}
\end{eqnarray}

\subsection{Major dimensionless parameters and their role}

The dimensionless form of equations (\ref{11psi})-(\ref{11g}) is given by 
\begin{equation}
L^{2}\frac{\partial ^{2}\Psi }{\partial Z^{2}}+R\frac{\partial }{\partial R}%
\left( \frac{1}{R}\frac{\partial \Psi }{\partial R}\right) =-R\Omega
_{\theta },  \label{12PSI}
\end{equation}
\begin{equation}
\frac{D\Omega _{\theta }/R}{DT}=-2K^{2}\frac{\Gamma \Omega _{r}}{R^{3}}+%
\frac{1}{\func{Re}}\left( \frac{1}{R}\frac{\partial }{\partial R}\left( 
\frac{1}{R}\frac{\partial \Omega _{\theta }R}{\partial R}\right) +L^{2}\frac{%
\partial ^{2}\Omega _{\theta }/R}{\partial Z^{2}}\right) ,  \label{12Wt}
\end{equation}
\begin{equation}
\frac{D\Gamma }{DT}=\frac{1}{\func{Re}}\left( R\frac{\partial }{\partial R}%
\left( \frac{1}{R}\frac{\partial \Gamma }{\partial R}\right) +L^{2}\frac{%
\partial ^{2}\Gamma }{\partial Z^{2}}\right) ,  \label{12G}
\end{equation}
where 
\begin{equation}
\frac{D}{DT}\equiv \func{St}\frac{\partial }{\partial T}+V_{z}\frac{\partial 
}{\partial Z}+V_{r}\frac{\partial }{\partial R},\;\;V_{z}=\frac{1}{R}\frac{%
\partial \Psi }{\partial R},\;\;V_{r}=-\frac{1}{R}\frac{\partial \Psi }{%
\partial Z},  \label{12DT}
\end{equation}
\begin{equation}
\Omega _{\theta }=L^{2}\frac{\partial V_{r}}{\partial Z}-\frac{\partial V_{z}%
}{\partial R},\quad \Omega _{r}=-\frac{1}{\func{St}}\frac{1}{R}\frac{%
\partial \Gamma }{\partial Z},\quad \Omega _{z}=\frac{1}{\func{St}}\frac{1}{R%
}\frac{\partial \Gamma }{\partial R}.  \label{12WW}
\end{equation}
The dimensionless parameters are introduced as 
\begin{equation}
\func{Re}\equiv L\frac{v_{\ast }r_{\ast }}{\nu },\;\;\func{St}\equiv \frac{%
r_{\ast }^{2}\omega _{\ast }}{\gamma _{\ast }}=\frac{r_{\ast }}{t_{\ast
}v_{\ast }L},\;\ K=\frac{(\gamma _{\ast }\omega _{\ast })^{1/2}}{v_{\ast }}%
,\ \;L\equiv \frac{r_{\ast }}{z_{\ast }}  \label{12par}
\end{equation}
and the variables are normalised according to 
\begin{equation*}
R=\frac{r}{r_{\ast }},\;Z=\frac{z}{z_{\ast }}\;\;\Psi =\frac{\psi }{\psi
_{\ast }},\;V_{r}=\frac{v_{r}}{v_{\ast }L},\;V_{z}=\frac{v_{z}}{v_{\ast }}%
,\;\Omega _{\theta }=\omega _{\theta }\frac{r_{\ast }}{v_{\ast }},
\end{equation*}
\begin{equation}
\Gamma =\frac{\gamma }{\gamma _{\ast }},\;\Omega _{z}=\frac{\omega _{z}}{%
\omega _{\ast }},\;\Omega _{r}=\frac{\omega _{r}}{\omega _{\ast }L},\;T=%
\frac{t}{t_{\ast }}\;.  \label{12nor}
\end{equation}
\emph{The subscript ''$\ast $''\ indicates constant characteristic values in
the region under consideration: $v_{\ast },$ $v_{\ast }L$ and $\gamma _{\ast
}/r_{\ast }$ represent the characteristic values of the axial, radial and
tangential velocity components, $\omega _{\ast }$ is the characteristic
axial vorticity and the parameter $L=r_{\ast }/z_{\ast },$ which is
generally considered to be of the order of unity here, specifies the
geometry of the region under consideration. The Reynolds number $\func{Re}$
determining the significance of viscous effects is typically very high in
vortical flows, while the Strouhal number $\func{St}$ characterises presence
of unsteady effects (Lundgren, 1985). The parameter $K$, which is called
here the \textit{vortical swirl ratio, }is discussed in the following
paragraphs.}

\emph{Note that not all of the characteristic values are independent: the
characteristic value of the stream function $\psi _{\ast }$ and the
characteristic time $t_{\ast }$ are determined by 
\begin{equation}
\psi _{\ast }=v_{\ast }r_{\ast }^{2},\;\;t_{\ast }=\frac{\gamma _{\ast }}{%
\omega _{\ast }v_{\ast }r_{\ast }L},
\end{equation}
while the parameters $r_{\ast },$ $z_{\ast },$ $v_{\ast }$, $\gamma _{\ast }$%
, $\omega _{\ast }$ and $\nu $ can be chosen freely. }The expression for the
characteristic time, $t_{\ast }$, is obtained from the convective terms of
equation (\ref{11g}): $\partial \gamma /\partial t\sim v_{r}\omega _{z}r$.%
\emph{\ The scale $\omega _{\ast }$ characterises axial vorticity $\omega
_{z}$ at $r=r_{\ast }$ which, generally, is located outside the viscous core
of the vortex.\ The problem under consideration is inherently unsteady if
axial vorticity $\omega _{z}$ is present in the surrounding flow. Indeed,
equation (\ref{11ww}) indicates that $\partial \gamma /\partial r=\omega
_{z}r>0$ when $\omega _{z}>0$ and, if viscous terms are neglected, equation (%
\ref{11g}) indicates that Lagrangian values of $\gamma $ are preserved.
Hence, as fluid flows towards the axis, the Eulerian value of $\gamma $ at a
given location must increase in time and the characteristic time of this
process $t_{\ast }$ is then controlled by (\ref{11ww}). This time
characterises the rate of circulation increase due to axial vorticity
present in the flow. }The small positive values of the Strouhal number $%
\func{St}$ indicate that the flow is close to its quasi-steady state,
although the flow is not exactly steady. As it shown by Lundgren (1985),
initially in a solid-body rotation $\func{St}\sim 1$ but as fluid particles
with high value of $\gamma $ move towards the axis, $\func{St}$ becomes
small (with exception of a rapidly shrinking region at the axis).

The parameter $K$ indicates the relative significance of \emph{axial}
vorticity present in the flow and \emph{controls the relative magnitude of
generation of the tangential vorticity as specified by equation (\ref{12Wt}%
). Note that presence of tangential vorticity in an axisymmetric flow
implies a helical structure of the vortex (investigated by Alekseenko et
al., 1999).} \ A detailed discussion of the role of this parameter is given
below. Other conventional dimensionless parameters --- the Rossby number $%
\func{Ro}$ and the swirl ratio $S$ --- can be expressed in terms of the
parameters introduced in (\ref{12par}) 
\begin{equation}
\func{Ro}\equiv \frac{v_{\ast }}{r_{\ast }\omega _{\ast }}=\frac{1}{K\func{St%
}^{1/2}},\ \ S\equiv \frac{\gamma _{\ast }/r_{\ast }}{v_{\ast }}=\frac{K}{%
\func{St}^{1/2}}\;.  \label{12parc}
\end{equation}
Note that this Rossby number is based on axial vorticity and not on the
Coriolis frequency (the former is typically much larger than the latter in
intensive vortices). The parameter $K=(S/\func{Ro})^{1/2}$ represents the
geometric mean of the conventional swirl ratio and the inverse Rossby
number. If rotation is close to a solid body rotation (i.e. $\gamma \approx
\omega _{z}r^{2}/2$) then there is little difference between these
parameters $K=S=1/\func{Ro}$. However, in other cases --- such as a
potential vortex ($\gamma \neq 0$, \ $\omega _{z}=0,$ $S\neq 0,$ $1/\func{Ro}%
=0$) --- the values of these parameters can be very different. The parameter 
$K$ takes a non-zero value only when both vorticity $\omega _{z}\neq 0$ and
circulation $\gamma \neq 0$ are present in the flow. It is the vortical
swirl ratio $K$ that determines the rate of generation of the tangential
vorticity $\Omega _{\theta }$ by equation (\ref{12Wt}).

If the parameter $K$ is small, the magnitude of the axial vorticity is
insufficient to generate any significant level of the tangential vorticity $%
\Omega _{\theta }$ by equation (\ref{12Wt}). This means that, in the regions
where the influence of boundary layers and buoyancy can be neglected, the $z$%
-$r$-image of the flow remains potential 
\begin{equation}
L^{2}\frac{\partial ^{2}\Psi }{\partial Z^{2}}+R\frac{\partial }{\partial R}%
\left( \frac{1}{R}\frac{\partial \Psi }{\partial R}\right) =-R\Omega
_{\theta }\approx 0\;.  \label{12PSI0}
\end{equation}
Complete description of this case is not difficult since vorticity is
passively transported by the flow.

The opposite case of very large values of $K$ ensures that the generated
tangential vorticity $\Omega _{\theta }$ is strong enough to significantly
affect the stream function $\Psi $ and the flow field. Which of the these
two limiting cases can better describe intensive vortical flows? In a
developed vortex, vorticity does affect the velocity components while the
assumption of weak vorticity and small $K$ results in a rather trivial
potential behaviour for the $r$-$z$ \emph{image of the flow} and is not
likely to be an acceptable model for the flow when the vortex is formed and
a noticeable level of axial vorticity is present in surrounding flow. The
case of strong vorticity and large $K$ seems much more relevant and is
considered in the following section.

\section{Strong vorticity in the intensification region}

This section considers a generic vortical flow with large values of the
vortical swirl ratio $K$. This ensures the presence of nonlinear
interactions between velocity and vorticity that play a significant role in
shaping the vortex. The vortex is generally presumed to be axisymmetric and
quasi-steady (with exception of Section \ref{sec_evol} and Appendix B where
unsteady effects are considered). A bathtub-like vortex is characterised by
fluid flowing towards the axis where the flow has a substantial axial
component. As fluid particles approach the axis, their rotation speed is
amplified and the region under consideration is called here the \textit{%
intensification region,} while vortices of this kind are called \textit{%
intense vortices}. \emph{This region is subject to axial stretch (which, as
shown in Appendix B, amplifies vorticity $\omega _{z}$) and is of prime
interest for our analysis. The intensification region is located away from
the layers with dominant influence of viscosity and the viscous terms are
neglected in Sections 3.2 and 3.3 (the details of the asymptotic treatment
of the viscous core is presented in Appendix A). This region is intermediate
between the viscous core (or aircore of a bathtub vortex) and the outer
flow, which can be represented by a sink-type flow but, generally, is
influenced by the surrounding conditions and becomes non-axisymmetric and
non-universal. } Consequently, the intensification region does not have its
own characteristic scale but is limited by the characteristic scales of the
viscous core and the outer flow. \emph{From the inner perspective, the
intensification region corresponds to the flow just outside the vortex core.
From the outer perspective, the intensification region is located in the
inner converging section of the outer flow where, in absence of a strong
swirl, the stream function would be approximated by the axisymmetric
converging flow $\psi \sim r^{2}z$.} The process of convective evolution of
vorticity dynamically coupled with the velocity field is of prime importance
to the intensification region. This section shows that, once a sufficiently
high value of $K$ is achieved, the velocity/vorticity interactions trigger a
compensating mechanism that limits variations of local value of the vortical
swirl ratio $K$ and play a certain stabilising role in the vortical flow.
While asymptotic expansions are based on large values of $K,$ increases of
the vortical swirl ratio trigger a compensating mechanism that moderates or
prevents further growth of this parameter, as discussed further in this
section.

\subsection{Strong swirl approximation\label{sec_sv}}

Different aspects of the strong swirl solution for axisymmetric flows were
introduced by Einstein and Li (1951), Lewellen (1962), Turner (1966),
Lundgren (1985) and Klimenko (2001 a-c). This approximation is characterised
by strong vorticity in the flow so that $1$/$K^{2}$ can be assumed to be
small. The dimensionless variables are represented in the form of the
following expansions 
\begin{equation*}
\Psi =\Psi _{0}+K^{-2}\Psi _{1}+...,\;V_{r}=V_{r0}+K^{-2}V_{r1}+...,\;\
V_{z}=V_{z0}+K^{-2}V_{z1}...,
\end{equation*}
\begin{equation*}
\Omega _{\theta }=\Omega _{\theta 0}+K^{-2}\Omega _{\theta
1}+...,,\;\;\Gamma =\Gamma _{0}+K^{-2}\Gamma _{1}+...,
\end{equation*}
\begin{equation}
\Omega _{r}=\Omega _{r0}+K^{-2}\Omega _{r1}+...,\;\;\Omega _{z}=\Omega
_{z0}+K^{-2}\Omega _{z1}+...  \label{A0ser}
\end{equation}
involving the higher-order terms as these are responsible for
vorticity/velocity interactions. Substitution of these expansions into
equations (\ref{12PSI})-(\ref{12WW}) results in 
\begin{equation}
V_{zi}=\frac{1}{R}\frac{\partial \Psi _{i}}{\partial R},\;\;V_{ri}=-\frac{1}{%
R}\frac{\partial \Psi _{i}}{\partial Z}\quad \Omega _{ri}=-\frac{1}{\func{St}%
}\frac{1}{R}\frac{\partial \Gamma _{i}}{\partial Z};\quad \Omega _{zi}=\frac{%
1}{\func{St}}\frac{1}{R}\frac{\partial \Gamma _{i}}{\partial R}
\label{A0def}
\end{equation}
\begin{equation}
R\Omega _{\theta i}=-R\frac{\partial }{\partial R}\left( \frac{1}{R}\frac{%
\partial \Psi _{i}}{\partial R}\right) -L^{2}\frac{\partial ^{2}\Psi _{i}}{%
\partial Z^{2}},\ \ \frac{D_{0}}{DT}\equiv \func{St}\frac{\partial }{%
\partial T}+V_{z0}\frac{\partial }{\partial Z}+V_{r0}\frac{\partial }{%
\partial R},\;
\end{equation}
\begin{equation}
\Omega _{r0}=0,\ \ \Gamma _{0}=\Gamma _{0}(R,T),\ V_{r0}=0,\ \ \ \Psi
_{0}=F_{0}\left( R,T\right) +F_{1}(R,T)\,Z,  \label{A0PHI}
\end{equation}
\begin{equation}
\func{St}\frac{\partial \Gamma _{0}}{\partial T}+V_{r0}\frac{\partial \Gamma
_{0}}{\partial R}=\frac{1}{\func{Re}}\left( R\frac{\partial }{\partial R}%
\left( \frac{1}{R}\frac{\partial \Gamma _{0}}{\partial R}\right) \right) ,
\end{equation}
\begin{equation}
2\frac{\Gamma _{0}}{R^{3}}\Omega _{r1}=-\frac{D_{0}\Omega _{\theta 0}/R}{DT}+%
\frac{1}{\func{Re}}\left( \frac{1}{R}\frac{\partial }{\partial R}\left( 
\frac{1}{R}\frac{\partial \Omega _{\theta 0}R}{\partial R}\right) \right) ,
\end{equation}
\begin{equation}
V_{r1}\frac{\partial \Gamma _{0}}{\partial R}=-\frac{D_{0}\Gamma _{1}}{DT}+%
\frac{1}{\func{Re}}\left( R\frac{\partial }{\partial R}\left( \frac{1}{R}%
\frac{\partial \Gamma _{1}}{\partial R}\right) +L^{2}\frac{\partial
^{2}\Gamma _{1}}{\partial Z^{2}}\right) ,  \label{A0G1}
\end{equation}
where $i=0,1$ and $F_{0}$ and $F_{1}$ are arbitrary functions determined by
the boundary conditions. \emph{It is easy to see that the leading-order
expressions in (\ref{A0PHI}) correspond to the conventional strong swirl
approximation}. Note that $V_{r}$ is only independent of $R$ at the leading
order. The terms of higher order indicate deviations from the leading-order
streamfunction (\ref{A0PHI}) induced by the strong vorticity/velocity
interactions.

The relatively slow rate of evolution that is common for intensive vortical
flows can be mathematically expressed by the condition $\func{St}\ll 1$. The
quasi-steady version of the strong swirl approximation is obtained with the
use of expansions 
\begin{equation*}
\Psi _{i}=\Psi _{i0}+\func{St}\Psi _{i1}+...,\;V_{ri}=V_{ri0}+\func{St}%
V_{ri1}+...,\;\ V_{zi}=V_{zi0}+\func{St}V_{zi1}...,
\end{equation*}
\begin{equation*}
\Omega _{\theta i}=\Omega _{\theta i0}+\func{St}\Omega _{\theta
i1}+...,\;\;\Gamma _{i}=\Gamma _{i0}+\func{St}\Gamma _{i1}+...,
\end{equation*}
\begin{equation}
\Omega _{ri}=\frac{1}{\func{St}}\Omega _{ri0}+\Omega _{ri1}+...,\;\;\Omega
_{zi}=\frac{1}{\func{St}}\Omega _{zi0}+\Omega _{zi1}+...,\ \ i=0,1\;.
\label{AAser}
\end{equation}
Several terms in these expansions (specifically $\Psi _{01}$,$\ \Gamma _{10}$
and the corresponding dependent terms $V_{z01}$,$\ \Omega _{z10}$, etc.) 
\emph{are not induced by the leading-order terms} and can be set to zero.
The leading and following-order equations are given by 
\begin{equation}
V_{zij}=\frac{1}{R}\frac{\partial \Psi _{ij}}{\partial R},\;\;V_{rij}=-\frac{%
1}{R}\frac{\partial \Psi _{ij}}{\partial Z}\quad \Omega _{rij}=-\frac{1}{R}%
\frac{\partial \Gamma _{ij}}{\partial Z},\quad \Omega _{zij}=\frac{1}{R}%
\frac{\partial \Gamma _{ij}}{\partial R},  \label{AA0def}
\end{equation}
\begin{equation}
R\Omega _{\theta ij}=-R\frac{\partial }{\partial R}\left( \frac{1}{R}\frac{%
\partial \Psi _{ij}}{\partial R}\right) -L^{2}\frac{\partial ^{2}\Psi _{ij}}{%
\partial Z^{2}},\ \ \frac{D_{00}}{DT}\equiv V_{z00}\frac{\partial }{\partial
Z}+V_{r00}\frac{\partial }{\partial R},\;
\end{equation}
\begin{equation}
\Gamma _{0i}=\Gamma _{0i}(R,T),\ \Omega _{r0i}=0,\ \ \Psi _{0i}=F_{0i}\left(
R,T\right) +F_{1i}(R,T)\,Z,
\end{equation}
\begin{equation}
\Psi _{01}=0,\ \ \Gamma _{10}=0,\ V_{r10}\frac{\partial \Gamma _{00}}{%
\partial R}=0,
\end{equation}
\begin{equation}
V_{r00}\frac{\partial \Gamma _{00}}{\partial R}=\frac{1}{\func{Re}}\left( R%
\frac{\partial }{\partial R}\left( \frac{1}{R}\frac{\partial \Gamma _{00}}{%
\partial R}\right) \right) ,\ \ \frac{\partial \Gamma _{00}}{\partial T}%
+V_{r00}\frac{\partial \Gamma _{01}}{\partial R}=\frac{1}{\func{Re}}\left( R%
\frac{\partial }{\partial R}\left( \frac{1}{R}\frac{\partial \Gamma _{01}}{%
\partial R}\right) \right) ,
\end{equation}
\begin{equation}
\ 2\frac{\Gamma _{00}}{R^{3}}\Omega _{r11}=-\frac{D_{00}\Omega _{\theta 00}/R%
}{DT}+\frac{1}{\func{Re}}\left( \frac{1}{R}\frac{\partial }{\partial R}%
\left( \frac{1}{R}\frac{\partial \Omega _{\theta 00}R}{\partial R}\right)
\right) ,
\end{equation}
\begin{equation}
V_{r11}\frac{\partial \Gamma _{00}}{\partial R}+V_{r10}\frac{\partial \Gamma
_{01}}{\partial R}=-\frac{D_{00}\Gamma _{11}}{DT}+\frac{1}{\func{Re}}\left( R%
\frac{\partial }{\partial R}\left( \frac{1}{R}\frac{\partial \Gamma _{11}}{%
\partial R}\right) +L^{2}\frac{\partial ^{2}\Gamma _{11}}{\partial Z^{2}}%
\right) ,  \label{AA0G1}
\end{equation}
where $i,j=0,1$.

\subsection{Inviscid solution in the intensification region}

As discussed previously, the convective evolution of vorticity is presumed
to be of prime importance for the intensification region. The inviscid
approximation of the quasi-steady strong vortex is now considered to obtain
a solution for the flow in the intensification region.\ One can put $\func{Re%
}^{-1}=0$ and simplify equations (\ref{AA0def})-(\ref{AA0G1}) 
\begin{equation}
\Gamma _{00}=\Gamma _{00}(T),\ \Gamma _{01}=\Gamma _{01}(R,T),\ \Omega
_{z00}=\Omega _{r00}=\Omega _{r01}=0,\ \Gamma _{10}=0,\   \label{ap_gam00}
\end{equation}
\begin{equation}
\Psi _{00}=F_{00}\left( R,T\right) +F_{10}(R,T)\,Z,\ \ \Psi _{01}=0,
\end{equation}
\begin{equation}
\frac{\partial \Gamma _{00}}{\partial T}=-V_{r00}\frac{\partial \Gamma _{01}%
}{\partial R}=-V_{r00}\Omega _{z01}R,
\end{equation}
\begin{equation}
2\frac{\Gamma _{00}\Omega _{r11}}{R^{3}}=-\frac{D_{00}\Omega _{\theta 00}/R}{%
DT},
\end{equation}
\begin{equation}
\ V_{r10}\frac{\partial \Gamma _{01}}{\partial R}=V_{r10}\Omega _{z01}R=-%
\frac{D_{00}\Gamma _{11}}{DT}\;.
\end{equation}
In an intensive vortical flow $F_{00}=0$ since $Z=0$ is a streamline. \emph{%
The behaviour of the vortical flows of this kind (for example, the Burgers
vortex) is conventionally examined in terms of radial power laws.} If the
stream is represented by a power law $F_{10}\sim R^{\alpha }$ with the
exponent $\alpha $ unknown a priori, the following consistent expressions,
which are analogous to the expressions obtained by Klimenko (2001b), are
recovered: 
\begin{equation}
\Psi _{00}=C_{0}R^{\alpha }Z,\;\;V_{r00}=-C_{0}R^{\alpha
-1},\;V_{z00}=\alpha C_{0}R^{\alpha -2}Z,  \label{ap_psi0}
\end{equation}
\begin{equation}
\Omega _{\theta 00}=-\alpha (\alpha -2)C_{0}R^{\alpha -3}Z,\;\;\Omega _{z01}=%
\frac{1}{R}\frac{\partial \Gamma _{01}}{\partial R}=-\frac{\Gamma
_{00}^{\prime }}{RV_{r0}}=\frac{\Gamma _{00}^{\prime }}{C_{0}R^{\alpha }},
\label{ap_gam0}
\end{equation}
\begin{equation}
\Gamma _{01}=-\frac{\Gamma _{00}^{\prime }}{(\alpha -2)C_{0}R^{\alpha -2}}%
,\;\;\Gamma _{00}^{\prime }\equiv \frac{\partial \Gamma _{00}}{\partial T},
\label{ap_gam1}
\end{equation}
\begin{equation}
\Omega _{r11}=2\alpha (\alpha -2)\frac{C_{0}R^{2\alpha -3}Z}{\Gamma _{00}}%
,\;\;\Gamma _{11}=-\alpha (\alpha -2)\frac{C_{0}^{2}R^{2\alpha -2}Z^{2}}{%
\Gamma _{00}},
\end{equation}
\begin{equation}
V_{r10}=2\alpha (\alpha -2)\frac{C_{0}^{4}R^{4\alpha -5}Z^{2}}{\Gamma
_{00}\Gamma _{00}^{\prime }},\;\;\Psi _{10}=\frac{2}{3}\alpha (\alpha -2)%
\frac{C_{0}^{4}R^{4\alpha -4}Z^{3}}{\Gamma _{00}\Gamma _{00}^{\prime }}.
\label{ap_psi1}
\end{equation}
The asymptotic correctness of the strong swirl approximation is determined
by the following parameter 
\begin{equation}
\varsigma \equiv \left| \frac{\Psi _{10}}{\Psi _{00}}\right| =\frac{2}{3}%
\alpha (\alpha -2)\frac{C_{0}^{3}Z^{2}}{\Gamma _{00}\Gamma _{00}^{\prime }}%
R^{3\alpha -4}\;.
\end{equation}
Large values of $\varsigma $ indicate that the asymptotic expansion
corresponding to the strong swirl approximation is no longer valid. If $%
\alpha <4/3$ and $\alpha \neq 0$ then $\varsigma \rightarrow \infty $ as $%
R\rightarrow 0.$ Hence, a strong swirl would not form and cannot be
sustained, if $\alpha $ is noticeably less than 4/3 over a wide range of
radii. The physical explanation for this fact is given in the next
subsections, \emph{where the implications of the power-law solutions are
analysed and the value of }$\alpha $\emph{\ is determined}.

Note that $\alpha =0$ and $\alpha =2$ represent special cases (vortical sink
and Burgers-type vortex) where the flow image on $r$-$z$-plane is potential
and the correcting terms are nullified $\Omega _{r11}=\Gamma
_{11}=V_{r10}=\Psi _{10}=0.$ A large $K$ is not needed to sustain the flow
in this case. Equations (\ref{ap_gam00})-(\ref{ap_psi1}) are formally valid
for $\alpha =0$ but the case of $\alpha =2$, which is more interesting for
the present study, requires special treatment: 
\begin{equation}
\Psi _{00}=C_{0}R^{2}Z,\;V_{r00}=-C_{0}R,\;V_{z00}=2C_{0}Z,\;\Omega _{z01}=%
\frac{\Gamma _{00}^{\prime }}{C_{0}R^{2}},\;\Gamma _{01}=\frac{\Gamma
_{00}^{\prime }}{C_{0}}\ln (R)\;.  \label{ap_alf2}
\end{equation}

\subsection{Downstream relaxation to the power law}

The power law approximations (\ref{ap_psi0})---(\ref{ap_psi1}) obtained in
the previous subsection are now used to analyse the flow in the
intensification region. The dimensional form of the leading-order equations
is given by 
\begin{equation}
\psi =f(r)\left( z+bz^{3}+...\right) ,\ v_{r}=v_{r0}+v_{r1}+...\;\ ,
\label{21psi}
\end{equation}
\begin{equation}
v_{r0}=-\frac{f(r)}{r},\;\ v_{r1}=-3bz^{2}\frac{f(r)}{r},\;\ 
\label{21psiaaa}
\end{equation}
\begin{equation}
v_{z}=\frac{f^{\prime }(r)}{r}z+...,\;\;\omega _{\theta }=-\left( \frac{%
f^{\prime }(r)}{r}\right) ^{\prime }z+...,\ \   \label{21psiaa}
\end{equation}
\begin{equation}
\ \ \ \omega _{z}=\frac{1}{r}\frac{\partial \gamma _{1}}{\partial r}=\frac{%
-\gamma _{0}^{\prime }}{rv_{r}}+...=\frac{\gamma _{0}^{\prime }}{f(r)}%
+...,\;\;  \label{21g}
\end{equation}
\begin{equation}
\gamma =\gamma _{0}(t)+\gamma _{1}(t,r)+\gamma _{2}(t,r)+...,\;
\label{21gaa}
\end{equation}
\begin{equation}
\frac{\mathrm{d}_{0}(\omega _{\theta }/r)}{\mathrm{d}t}=-2\frac{\gamma
_{0}\omega _{r}}{r^{3}}+...,\;\;\;\;\omega _{r}=-2\frac{\gamma _{2}}{rz}+...,
\label{21e2}
\end{equation}
\begin{equation}
\frac{\mathrm{d}_{0}\gamma _{2}}{\mathrm{d}t}=-v_{r1}\omega _{z}r+...,\;\;\;%
\frac{\mathrm{d}_{0}}{\mathrm{d}t}\equiv v_{z}\frac{\partial }{\partial z}%
+v_{r0}\frac{\partial }{\partial r}\;.  \label{21e1}
\end{equation}
Here, $\gamma _{0}^{\prime }=\mathrm{d}\gamma _{0}/\mathrm{d}t$ is
introduced and the higher-order terms that are needed to form a consistent
link between velocity and vorticity are retained. Note that substituting the
streamfunction in the power-law form 
\begin{equation}
\psi =c_{0}r^{\alpha }z+...,\;\;f(r)=c_{0}r^{\alpha }  \label{plaw}
\end{equation}
into equations (\ref{21psi})-(\ref{21e1}) results in 
\begin{equation}
\alpha =\alpha ^{\ast }=\frac{4}{3},\;\;c_{0}^{3}=\frac{27}{16}b\gamma
_{0}\gamma _{0}^{\prime }\   \label{21c0}
\end{equation}
and in the previously obtained asymptotic equations (\ref{ap_psi0})-(\ref
{ap_psi1}). The value $\alpha ^{\ast }=4/3$ is called the compensating value
of the exponent $\alpha $. For the power law, the horizontal flow
convergence $\lambda $ (which is the same as the axial stretch) \ and
tangential velocity $v_{\theta }$ are determined by the equations 
\begin{equation}
\lambda =-\frac{1}{r}\frac{\partial v_{r}r}{\partial r}=\frac{\partial v_{z}%
}{\partial z}=c_{0}\alpha r^{\alpha -2}+...,  \label{lam}
\end{equation}

\begin{equation}
\omega _{z}=\frac{\gamma _{0}^{\prime }}{c_{0}r^{\alpha }}\Longrightarrow
v_{\theta }=\frac{\gamma }{r}=\frac{\gamma _{0}}{r}+\frac{\gamma _{1}}{r}%
+...,\;\;\frac{\gamma _{1}}{r}=\frac{\gamma _{0}^{\prime }}{c_{0}(2-\alpha
)r^{\alpha -1}}\;.  \label{12vt}
\end{equation}

The behaviour of the vortex under conditions, in which the structure of the
flow is preserved but $f(r)$ deviates from the power law, is now
investigated. Asymptotic solutions obtained by Klimenko (2001c) indicate
that vortical flows behave differently depending on whether disturbances,
which are introduced into the flow, are gradual or sudden. Gradual
disturbances preserve the structure of the strong swirl approximation while
sudden disturbances tend to violate this approximation and produce
propagating waves. Here, we are interested only in gradual changes and,
according to the method of multi-scale expansions, seek a solution in the
form $f=c(\xi )\tilde{r}^{\alpha },\;$where $\xi =\ln (r)$ is treated as the
slow variable and $\tilde{r}=r$ is treated as the fast variable. The
derivatives are now expressed by 
\begin{equation*}
\frac{\mathrm{d}}{\mathrm{d}r}=\frac{\mathrm{d}}{\mathrm{d}\tilde{r}}+\frac{1%
}{\tilde{r}}\frac{\mathrm{d}}{\mathrm{d}\xi }\;.
\end{equation*}
In the following equations, all derivatives with respect to $\tilde{r}$ are
retained in the equations, but only $c(\xi )$ and its first derivative with
respect to $\xi $ (i.e. the leading- and the next-order terms) are
considered. At the leading order, equations (\ref{21c0}) are obtained. The
governing equations at the next order take the form 
\begin{equation}
\omega _{\theta }=\frac{z}{r^{3-\alpha }}\left( \alpha (2-\alpha )c-2(\alpha
-1)\frac{\mathrm{d}c}{\mathrm{d}\xi }+...\right) ,
\end{equation}
\begin{equation}
\gamma _{2}=\frac{z^{2}r^{2\alpha -2}}{\gamma _{0}}\left( \alpha (2-\alpha
)c^{2}-2(\alpha -1)c\frac{\mathrm{d}c}{\mathrm{d}\xi }+...\right) ,
\label{21e3a}
\end{equation}
\begin{equation}
\frac{r^{3\alpha -4}}{\gamma _{0}}\left( 2\alpha (2-\alpha )c^{3}-4(\alpha
-1)c^{2}\frac{\mathrm{d}c}{\mathrm{d}\xi }+...\right) =3b\gamma _{0}^{\prime
}\;.  \label{21e3}
\end{equation}
Here, $\omega _{\theta }$ is determined from equation (\ref{21psiaa}), $%
\gamma _{2}$ is expressed in terms of $\omega _{r}$, which in turn is
determined by $\omega _{\theta }$ according to (\ref{21e2}), while the last
equation is obtained from (\ref{21e1}) \emph{\ with $\gamma _{2}$ given by (%
\ref{21e3a}). \ Finally, equation (\ref{21e3}) can be written in the form of
downstream relaxation 
\begin{equation}
\frac{\mathrm{d}c}{\mathrm{d}(-\xi )}=\frac{4}{3}\left( \frac{c_{0}^{3}}{%
c^{2}}-c\right) \;.  \label{21e4}
\end{equation}
This equation indicates that, as $\xi $ decreases, $c(\xi )$ relaxes
downstream\ to its constant equilibrium value $c_{0}$ given by (\ref{21c0})
. The solution of this equation $c^{3}=c_{0}^{3}+k\exp \left( 4\xi \right) ,$
where $k$ is constant$,$ yields } 
\begin{equation}
f(r)=c_{0}r^{\alpha ^{\ast }}\left( 1+\frac{k}{3c_{0}^{3}}r^{4}+...\right)
\;.  \label{21e5}
\end{equation}
It is easy to see that after a deviation of the function $f$ from the radial
power-law, $f$ tends to return downstream back to the compensating power law
as $c(r)$ relaxes to $c_{0}$. \emph{\ Note that, as specified by (\ref{21c0}%
), $\alpha ^{\ast }=4/3$ in equation (\ref{21e5}) and that the relaxation
mechanism is valid only for centripetal but not for centrifugal direction of
the flow. The following subsection offers a qualitative explanation of this
effect and shows that in realistic vortical flows the compensating exponent
needs to be extended from the single value of $\alpha ^{\ast }=4/3$ to the
narrow range of $4/3\leq \alpha ^{\ast }\leq 3/2$. }

\subsection{The compensating mechanism in the vortical flow}

\begin{figure}
\centering
\includegraphics[width=11cm,trim=3cm 12cm 4cm 2cm,clip]{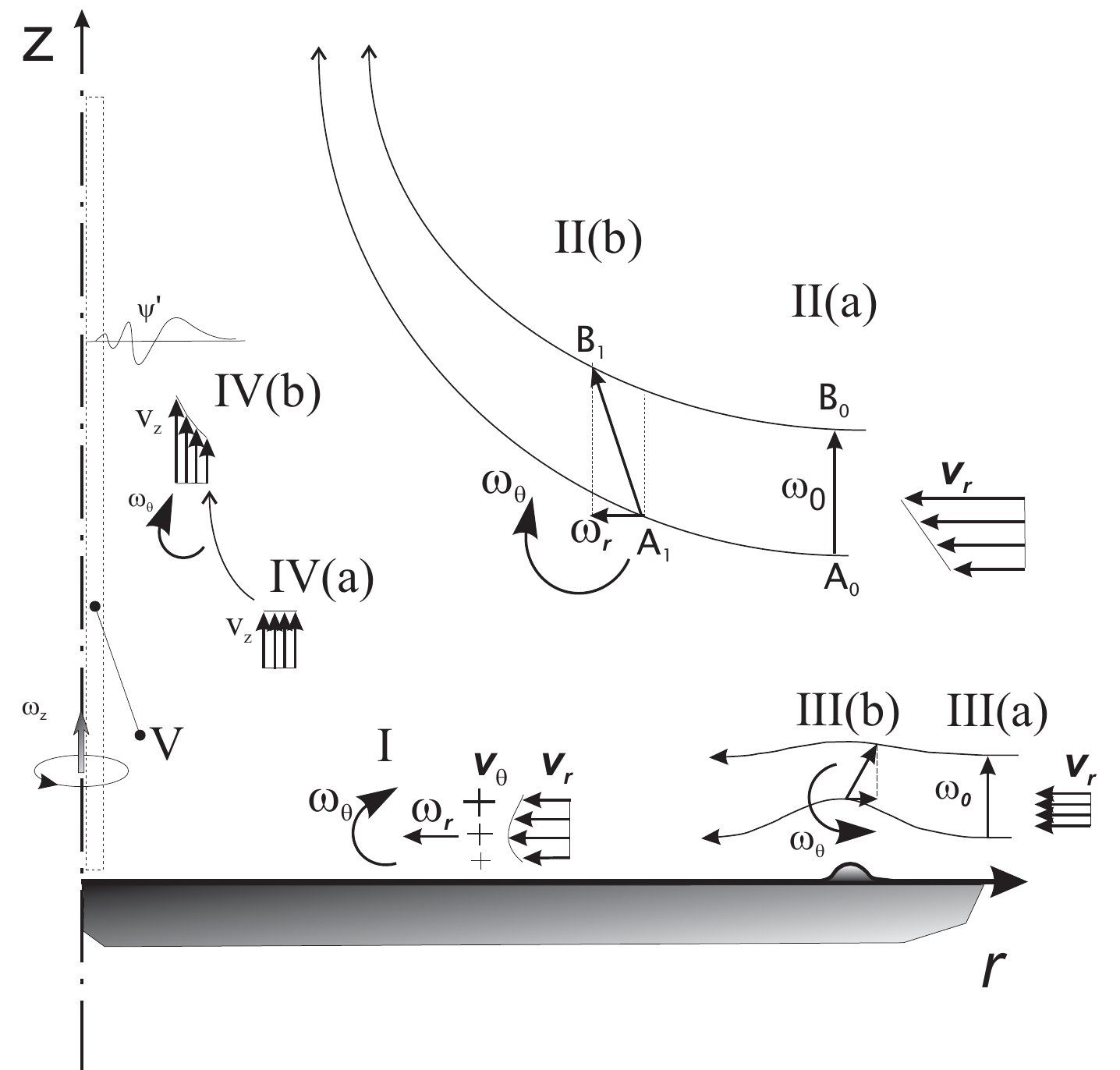} 
\caption{Schematic of vorticity evolution in intensive vortical
flows.}
\label{fig1} 
\end{figure}

While interactions between velocity and vorticity are commonly a
destabilising factor in fluid flows, the axisymmetric vortical flows
considered here have a certain stabilising mechanism linked to evolution of
vorticity. If a disturbance is introduced into these flows, the generated
tangential vorticity $\omega _{\theta }$ tends to compensate for this
disturbance and preserve the overall structure of the flow. This effect is
illustrated in figure \ref{fig1} where case III shows the flow over an
axisymmetric small disturbance and vorticity $\omega _{\theta }$ is
generated to compensate for the disturbance and preserve the independence of 
$\gamma $ from $z$ at the leading order. More detailed explanations of the
stabilising evolution of vorticity and a full asymptotic solution for this
problem are given in Klimenko (2001c). A similar mechanism, which acts to
compensate for deviations from the power-law (\ref{21c0}), is analysed in
this subsection.

Figure \ref{fig1} also illustrates the direction of vorticity $\omega
_{\theta }$ which tends to be generated in the vortical flows of this
geometry. There are several effects, both inviscid and viscous, that are
responsible for the presence of negative $\omega _{r}$ generating $\omega
_{\theta }$ according to equation (\ref{11wt}). The first effect is,
essentially, the Ekman effect. If \ $v_{\theta }=0$ at the lower boundary
this corresponds to negative vorticity $\omega _{r}$ that generates
vorticity $\omega _{\theta }$ in the direction shown in figure \ref{fig1}
(case I). Another effect appears due to the existence of vertical shear in a
typical profile of $v_{r}$ (case II in figure \ref{fig1}). This shear may
appear due to no-slip conditions at the lower boundary or due to inviscid
effects in a bathtub flow (see Klimenko, 2001a for details). Consistency
between the strong vortex and the boundary layer induced by the no-slip
conditions acts as a factor constraining the flow (Turner, 1966).

Assuming that the vorticity vector is frozen into the flow (as illustrated
by the joint evolution of the vorticity and material vectors transforming $%
\overrightarrow{\text{A}_{0}\text{B}_{0}}$ into $\overrightarrow{\text{A}_{1}%
\text{B}_{1}}$), this shear causes the presence of negative $\omega _{r}$ at
location IIb in figure \ref{fig1}. Faster rotation at smaller radii turns
the vorticity vector away from the reader, resulting in the appearance of $%
\omega _{\theta }$ in the direction shown.\ Although the exact convective
mechanisms of generating tangential vorticity $\omega _{\theta }$ may be
somewhat different for different vortices, generation of $\omega _{\theta }$
acting in the direction of lowering the values of exponent $\alpha $ below $%
2 $ and stimulating updraft (as illustrated by case IV in figure \ref{fig1}%
)\ is common for the intensive vortices.

Many vortical flows have a sufficiently wide range of radii to create
conditions for substantial amplification of axial vorticity. Since different
characteristic radii can be characterised by different characteristic values
of $K$, the localised version of the vortical swirl ratio is introduced and
defined in terms of local parameters by 
\begin{equation}
K_{r}^{2}=\frac{\gamma \omega _{z}}{v_{z}^{2}}\sim \frac{\gamma }{%
c^{3}r^{3\alpha -4}}\sim \frac{\gamma _{0}+\gamma _{1}}{c^{3}r^{3\alpha -4}}%
\;.  \label{21K}
\end{equation}
Here and in the rest of this subsection, equations (\ref{21psi})-(\ref{21e1}%
) with power-law approximation of the streamfunction $f(r)=cr^{\alpha }$ are
used. The subscript ''$r$'' indicates that $K_{r}$ is radius-dependent: in
principle, the local vortical swirl ratio $K_{r}$ may exhibit a strong
dependence on $r.$ If, for example, $\alpha =2,$ then $K_{r}\rightarrow
\infty $ as $r\rightarrow 0$. According to equation (\ref{ap_gam0}), $\alpha
=2$ corresponds to a flow with $\omega _{\theta }=0$ (i.e. with potential
image on the $r$-$z$-plane)$,$ while smaller values of $\alpha <2$ imply
higher values of vorticity $\omega _{\theta }$ and more rapid increase of $%
\omega _{\theta }$ towards the axis as $r\rightarrow 0$. The relative
magnitude of vorticity $\omega _{\theta }$ generated in the vortical flow is
determined by the local vortical swirl ratio $K_{r}$. The parameter $K_{r}$
cannot unrestrictedly \textit{decrease} towards the axis ($r\rightarrow 0$)
since small values of $K_{r}$ correspond to negligible $\omega _{\theta }$
and, consequently, to $\alpha \rightarrow 2,$ which results in $K_{r}$ 
\textit{increasing} towards the axis according to (\ref{21K}). At the same
time $K_{r}$ cannot unrestrictedly \textit{increase} towards the axis as
this would produce large quantities of $\omega _{\theta }$ that decrease the
effective value of $\alpha $ up until $K_{r}$ is forced by (\ref{21K}) to 
\textit{decrease} towards the axis. This mechanism compensating for changes
of $\alpha $ is reflected in equation (\ref{21e4}).

While excessively low values of $K_{r}$ render the strong swirl
approximation inapplicable, excessively large values of $K_{r}$ can cause
bifurcations or destabilise the flow. Theory for vortex breakdown in
steady-state axisymmetric inviscid flows governed by the Long-Squire
equation\ was introduced by Benjamin (1962) \emph{and applied to confined
vortices by Escudier et al. (1982).} Klimenko (2001b) found that Benjamin's
theory can be used near the axis of intensive vortical flows, where the
axial component of velocity is dominant (case IV in figure \ref{fig1}). The
Benjamin equation (which represents a perturbed version of the Long-Squire
equation) has a useful analytical solution for the power-law dependence of $%
K_{r}$ on $r$ (see Klimenko, 2001b). This solution indicates that, if $%
K_{r}\rightarrow \infty $ as $r\rightarrow 0,$ then vortex breakdown is
expected according to this solution.

\emph{It follows that, although $K_{r}$ can depend on $r$ according to the
definition of this parameter, this dependence should be weak when the swirl
is strong. Indeed, the condition $K_{r}\sim \func{const}$ ensures that the
tangential vorticity is neither overproduced to destabilise the flow nor
underproduced to result in the contradictions mentioned above. The regime
that corresponds to these conditions and compensates for possible increases
or decreases of $K_{r}$ is called here the \textit{compensating regime}. }%
The mechanism of vorticity/velocity interactions associated with this regime
co-balances $v_{z}^{2}$ with $\gamma \omega _{z}$ and counter-balances $%
\omega _{z}$ with $\gamma $ to keep $K_{r}\sim \func{const}$.

\emph{The compensating regime is linked to the compensating value of the
exponent $\alpha $.} Consider the two circulation terms retained in (\ref
{21K}): $r$-independent $\gamma _{0}$ and $r$-dependent $\gamma _{1},$ which
is linked to $\omega _{z}$ by (\ref{21g}) so that $\gamma _{1}\sim
r^{2-\alpha }.$ The relative magnitudes of the terms $\gamma _{0}$ and $%
\gamma _{1}$ in (\ref{21K}) combined with the condition $K_{r}\sim \func{%
const}$ \ determine different expressions for the compensating value $\alpha
^{\ast }$ of the exponent $\alpha $ 
\begin{equation}
\ 
\begin{array}{cc}
\alpha ^{\ast }=4/3, & \gamma _{0}\gg \gamma _{1} \\ 
\alpha ^{\ast }=3/2, & \gamma _{0}\ll \gamma _{1}
\end{array}
\label{21K1}
\end{equation}
Practically, this means the exponent $\alpha =\alpha ^{\ast }$ of the
compensating regime is not a fixed value but is represented by the range of\ 
$4/3\leq \alpha ^{\ast }\leq 3/2.$ If $\alpha <4/3$ over a wide range of
radii, the strong swirl cannot form at the axis. If $\alpha >3/2$ persists,
large $K_{r}$ near the axis would cause vortex breakdown followed by
weakening of the swirl. Unsteady effects are discussed in the next
subsection and in Appendix B.

The asymptotic analysis of the previous subsections requires that $\gamma
_{1}/\gamma _{0}\sim \func{St}\ll 1$ in expansion (\ref{AAser}) or,
otherwise, expansions (\ref{AAser}) would formally lose asymptotic
precision. The condition $\gamma _{1}/\gamma _{0}\ll 1$ corresponds to $%
\alpha ^{\ast }=4/3$ obtained in equation (\ref{21c0}), which neglects
losses of $\gamma _{0}$ and thus considers this term as dominant. Note that $%
\gamma _{0}\gg \gamma _{1}$ is not necessarily the case in realistic
vortices due to vortex breakdowns and loss of angular momentum into the
ground (which are not taken into account in the idealised considerations).
Since $\gamma _{1}$ increases with increasing $r$, the outer sections of the
intensification region are more likely to have a larger value of $\alpha
^{\ast }$ (within the compensating range) than the inner sections.

\emph{\ The compensating exponent is thus extended from the value $\alpha
^{\ast }=4/3$ to the range of $4/3\leq \alpha ^{\ast }\leq 3/2$. The
mechanism of evolution of vorticity in converging flows, which is reflected
by equation (\ref{21e4}) and explained above, acts to compensate for changes
of $K_{r}$. The downstream relaxation of $c(r)$ to its constant value $c_{0}$
governed by (\ref{21e4}) implies that $K_{r}$, which is defined by (\ref{21K}%
) and linked to the inverse third power of $c$, undergoes a similar
relaxation to its radius-independent value. }

\subsection{Evolution of the vortex \label{sec_evol}}

This subsection considers some of more complex transient effects in the
formation and breakdown of intensive vortices. Typically, the formation of
intensive vortices starts when a converging fluid motion occurs in the
presence of some background axial vorticity. One can assume that the
vorticity is initially distributed as in solid body rotation, $\omega
_{z}=\omega _{0}=\func{const}$. The initial value of the vortical swirl
ratio $K_{r}$ is small and rotation in the flow is not intense. The focus of
the present consideration is a near-axis region $r\leq r_{2}$ where the
stream function can be represented by $\psi =c_{0}r^{\alpha }z$. Since,
initially, $K_{r}$ is uniformly small in this region, the flow image on the $%
z$-$r$-plane is potential and $\alpha =2$. In this case $v_{z}=v_{0}=\func{%
const}$ at a given height $z=z_{0}$.\ This region is primarily responsible
for formation of the vortex. The other limiting case of $\alpha =0$
corresponds to two-dimensional flow called a vortical sink. This flow may be
relevant to more peripheral regions of the vortex where the value of $K_{r}$%
\ does not change significantly even without the presence of the
compensating mechanism. The inviscid solutions for evolution of vorticity
and circulation are presented in Appendix B.

\begin{figure}
 \centering
\includegraphics[width=8cm,trim=6cm 1.8cm 7cm 3cm,clip]{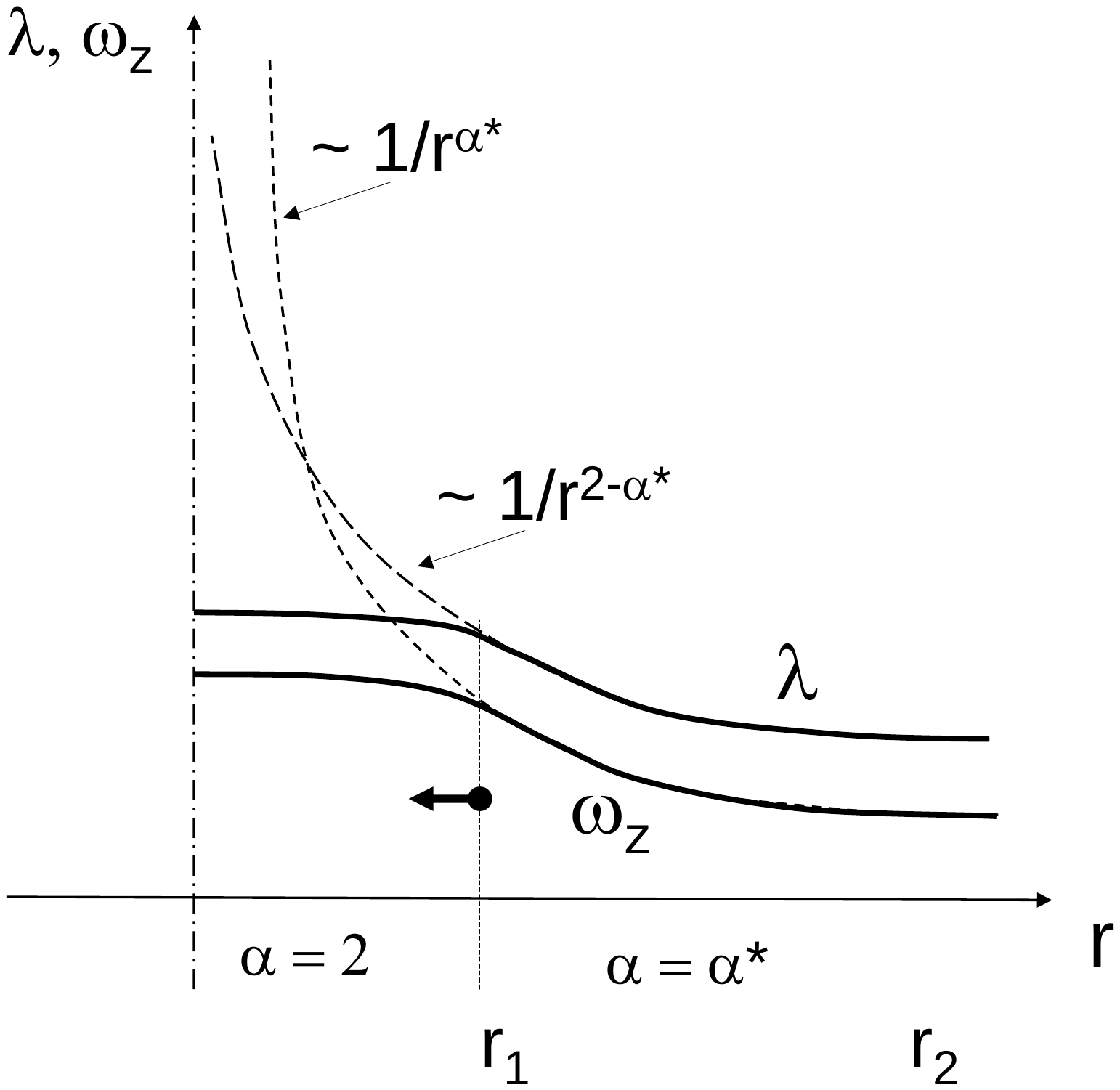} 
\caption{Schematic of development of the axial vorticity and horizontal convergence 
during formation of an intensive vortex; \emph{ current (------) and quasi-steady (-- -- --) distributions are shown.} }
\label{fig2}
\end{figure}

In the case of $\alpha =2,$ the value of $K_{r}$ is specified by the
following expression 
\begin{equation}
K_{r}^{2}=\frac{\gamma \omega _{z}}{v_{z}^{2}}=\exp (4\tau )\frac{\omega
_{0}^{2}}{2v_{0}^{2}}r^{2},  \label{22K}
\end{equation}
where $\tau $ is a time-like variable defined in Appendix B. The parameter $%
K_{r}$ rapidly increases with time and one can note that the quasi-steady
distribution of axial vorticity $\omega _{z}\sim 1/r^{\alpha }$ is never
achieved as long as $\alpha $ remains 2. The growth of $K_{r}$ can not
continue indefinitely and at a certain moment $K_{r}\sim 1$ is achieved at
the outer rim of the region under consideration where, say, $r=r_{1}=r_{2}$.
The value $r_{2}$\ representing the radius of the domain is kept constant,
while the radius $r_{1}(t)$ where $K_{r}\sim 1$ continues to decrease
according to equation (\ref{22K}). The nature of the flow in the ring $%
r_{1}<r<r_{2}$ changes so $\alpha $ decreases and becomes close to $\alpha
^{\ast }$ since large values of $K_{r}$ are attained there. Since $\alpha <2$
within the ring, the convergence of the flow increases towards the axis. The
equations presented in Appendix B show that the axial vorticity does
approach the quasi-steady distribution $\omega _{z}\sim 1/r^{\alpha }$ when $%
\alpha <2$. In the region $r<r_{1}(t)$ the value of $K_{r}$\ is small and
insufficient to change the nature of the flow, hence $\alpha =2$ remains
there. Figure \ref{fig2} gives a schematic illustration of the formation of
an intensive vortex. Both the stretch $\lambda $ and axial vorticity $\omega
_{z}$ evolve from initial constant values to quasi-steady exponents of the
compensating regime (see also Appendix B). Formation of the vortex is
completed when $r_{1}$ approaches the axis and becomes small. This formation
is characterised by $\alpha $ falling from 2 to $\alpha ^{\ast }$ in $\psi
\sim r^{\alpha }z$, while the axial vorticity, which is initially constant,
increases its radial slope to converge to $\omega _{z}\sim 1/r^{\alpha },$
where $\alpha $ reaching $4/3$ is needed for a strong swirl to appear near
the axis.

During formation of the vortex, the flow undergoes a significant change.
Initially, when $K_{r}\ll 1$, the vorticity is present in the flow as a
passive quantity that is transported by the flow but does not significantly
affect the velocity field. At this stage the flow image on the $r$-$z$ plane
can be treated as potential since the vorticity level is low. A potential
flow immediately (with the speed of sound) reacts to any disturbance at the
boundaries of the domain under consideration and specifying the velocity
field at some imaginary boundaries of the flow uniquely determines the flow
within the region. The convergence point (which is selected as the origin of
coordinates $r=0$ in the axisymmetric model) is also fully determined by the
conditions on the imaginary boundary somewhere in the peripheral region of
the flow. In practice, this means that the convergence point would move
promptly and randomly if random disturbances are present at the periphery of
the flow, or the flow may even have several convergence points at a given
moment.

When the vortical swirl ratio $K_{r}$ becomes sufficiently large, a
noticeable amount of tangential vorticity $\omega _{\theta }$ is generated.
The structure of the flow changes, producing higher convergence near the
axis and compensating further increases in vorticity to keep $K_{r}$
constant. In this case, as discussed previously, vorticity exhibits some
stabilising effect on the flow. When the vortex is formed, the stabilising
effect of vorticity propagates towards the axis as $r_{1}$ decreases. The
tangential vorticity $\omega _{\theta }$ stimulates updraft near the flow
axis and vorticity evolves in time but does not respond immediately to
fluctuations at the boundaries of the region under consideration. This makes
the position of the centre of convergence, which is also the centre of the
vortical motion in our model, more stable. In this vortex, rotation is
intense and the vortex is now fully perceived as an intensive vortex.

Qualitative evolution of vorticity during the formation of a strong vortex
is shown in figure \ref{fig2}. Initially, the value of $\gamma _{0}$ defined
in (\ref{21g}) can be negative but, as the vortex forms, both $\omega _{z}$
and $\gamma _{0}$ increase. Without losses, the value of $\gamma _{0}$
continues to grow slowly but unrestrictedly; although, practically, any
substantially positive value of $\gamma _{0}$ would induce high velocities
near the axis $v_{\theta }\sim r^{-1},$ amplifying the losses of angular
momentum in the vicinity of the vortex core. A typical intensive vortex
remains stable and seems to be nearly stationary. The state of the vortex
flow, however, is determined by two major effects that control the relative
magnitude of $\gamma _{0}$ in (\ref{21g}): the increase of angular momentum
due to inflow from peripheral regions and the loss of angular momentum into
the physical boundaries of the flow. If the influx of momentum exceeds its
losses, $\gamma _{0}$ continues to grow. While changes in the balance of
influx and losses may also disturb the vorticity profile, the equilibrium
state for this profile is specified by the exponents of the compensating
regime. As previously discussed, relatively large values of $\gamma _{0}$
are sustainable when $\alpha ^{\ast }=4/3$.

A persistent growth in $\gamma _{0}$ can increase $K_{r}$ to the level when
the flow bifurcates (as discussed in the previous subsection). \emph{The
analysis of the viscous core in Appendix A indicates that the singularity of 
$\alpha =\alpha ^{\ast }$ in $\psi \sim r^{\alpha }z$ disappears near the
axis and the value of $K_{r}$ is higher in the core than in the surrounding
inviscid flow; hence the bifurcations are likely to appear first in or near
the core} (schematically shown as region V in figure \ref{fig1}). This
effect, which causes reversal of the flow at the axis (i.e. from updraft to
downdraft), is known as vortex breakdown (see, for example, \emph{Escudier
et al., 1982;} Lewellen, 1993; Church and Snow, 1993; Lee and Wurman, 2005).
Axial downdrafts are common during late stages of development of very
intensive vortices (Church and Snow, 1993; Emanuel, 2003). Vortex breakdowns
affect the structures of the cores of tornadoes and the eyes of hurricanes
by terminating the centripetal flow near the axis and thus reducing the
relative value of $\gamma _{0}$. The balanced value of $\alpha ^{\ast }$
that corresponds to negligible $\gamma _{0}$ is $3/2.$

The vorticity equations allow for an alternative scheme of vortex formation.
Let us assume that the initial $K_{r}$ is small and that $\omega _{z}$ is
initially distributed in a quasi-steady manner $\omega _{z}\sim 1/r^{\alpha
} $. Since $K_{r}$ is small, $\alpha =2.$ Under quasi-steady conditions, the
vortical swirl ratio is given by equation (\ref{21K}) $K_{r}\sim \gamma
/r^{2}$. As $\gamma $ increases due to additional angular momentum brought
from peripheral regions, $K_{r}$ becomes large first near the axis and the
radius $r_{1}(t)$ of $K_{r}\sim 1$ then moves sideways. The compensating
regime appears first near the axis $\alpha =\alpha ^{\ast }$ at $r<r_{1}$
and propagates sideways (where $\alpha =2$ at $r>r_{1}$). This scheme was
visualised by Klimenko (2001a) as a series of quasi-steady pictures of
calculated streamlines. This scheme, although possible, raises a question:
why is the quasi-steady solution reached in the first place under conditions
when $\alpha =2$ while, according Appendix B, $\omega _{z}$ approaches the
quasi-steady solution only when $\alpha <2$? Note that $\psi \sim r^{\alpha
}z$ is only the leading term in representation of the stream function and
the solution $\omega _{z}=\omega _{0}\exp (2\tau )$ may, in principle, be
altered by the other terms.

The formed intensive vortex is quite stable but does not persist forever. If
vorticity with a dominant direction is initially present in the tub, the
bathtub vortex is likely to persist until the tub is emptied, but intensive
vortices in the atmosphere can weaken and disappear for various reasons. The
influx of axial vorticity can become exhausted at a certain moment (due to
weakening of the recirculating motion or insufficient axial vorticity level
in the peripheral regions) but this does not mean that the vortex disappears
immediately. A significant amount of axial vorticity can be concentrated in
the viscous core (see Appendix A) and will maintain visible rotation even if 
$\omega _{z}=0$ in the surrounding flow. This case corresponds to $\alpha =2$
in $\psi \sim r^{\alpha }z$ since $K_{r}=0$ outside the core. At this stage,
the vortex can be characterised by the conventional stationary axisymmetric
solution of Lewellen (1962). Practically, the vortex would continue to lose
angular momentum to the physical boundaries of the flow until it fully
disappears.

\section{Atmospheric vortices}

Examples of intensive vortices of different scales are considered in this
section. The smallest vortex can be observed in a bathtub flow while the
largest are represented by atmospheric vortices -- dustdevils, firewhirls,
tornadoes, mesocyclones and cyclones. For bathtub flows, the compensating
exponents have been observed in computations (Klimenko, 2001a; Rojas, 2002)
and experiments (Klimenko, 2007; see also Shiraishi and Sato, 1994). The
atmospheric vortices are quite different from the bathtub vortex and from
each other. These differences stem from differences in scales and the
physical mechanisms that are responsible for the formation of these
vortices. There are, however, features that are common: the vortices can
usually be characterised by certain inner (core) scales and are embedded
into outer flows. As discussed in the previous sections, the present work
does not seek a complete description of these vortices; rather it neglects
less significant details (for example, rain bands of a hurricane) and
focuses on the main features of the average flow in the intensification
region, which, from the asymptotic perspective, is an intermediate region
between the inner and outer scales of the vortex. This region is
characterised by the presence of the axial vorticity and its continuing
amplification. Due to the reduced number of parameters needed to
characterise the intensification region, vortices of different scales may
exhibit a greater degree of similarity within this region. Among atmospheric
vortices, cyclones, tornadoes and, on some occasions, firewhirls are
characterised by an extended range of scales, greater stability and
resistance to atmospheric fluctuations (due to the enormous scales of
tropical cyclones, the high wind speeds achieved in tornadic flows, and the
extreme energies released by fires). This, of course does, not mean that
these vortices are completely regular -- the flow patterns in cyclones and
especially in tornadoes reflect irregularities of surrounding atmospheric
motions and display significant variation of flow parameters.

\emph{Tangential winds, which possess a significant inertia and are the
strongest in atmospheric vortices, are commonly reported and discussed in
publications. Tangential winds tend to be least affected by the surface
boundary layer and outer disturbances always present in the atmosphere.
Equations (\ref{12vt}) indicate that the tangential velocity and axial
vorticity of intensive vortices can be approximated by the following power
laws 
\begin{equation}
v_{\theta }=\frac{\gamma }{r}=\frac{\gamma _{0}}{r}+\frac{c_{1}}{(2-\alpha
)r^{\beta }}+...,\;\;\omega _{z}=\frac{1}{r}\frac{\partial \gamma }{\partial
r}=\frac{c_{1}}{r^{\alpha }}+...,\;\;\beta =\alpha -1  \label{alf_bet}
\end{equation}
and checked against experimental measurements. Equations (\ref{alf_bet})
have a two-term expression for $v_{\theta },$\ while the corresponding
approximation for $\omega _{z}$\ involves only one term.\ In many realistic
vortices, $\gamma _{0}$\ is small compared to the second term and either can
be neglected or can not be reliably determined from data available for a
limited range of radii. When $\gamma _{0}$\ \ is substantially positive,
attempting to fit the tangential velocity data by a single term $v_{\theta
}\sim 1/r^{\beta }$\ would result in overestimating $\beta $. Hence,
comparison of the theory and measurements based on vorticity is more direct
but, if significant noise is present in the data, numerical differentiation
of velocity profile can produce high levels of scattering.}

\subsection{Firewhirls}

Firewhirls are fires that are characterised by the presence of a strong
rotation in the flow resulting in elongated and more intense flames
(Williams, 1982). Once firewhirls appear in a fire, they greatly intensify
burning and are very difficult to extinguish. The scales of firewhirls are
generally comparable to those of tornadoes, which are discussed in the
following subsection. In firewhirls, however, the centripetal flow is
stimulated by a large heat release and buoyant uplifting, which should
prevent the vortex breakdowns and axial downdrafts typical in other
intensive atmospheric vortices. Firewhirls observed on inclined surfaces are
most interesting as they can deviate from the vertical direction and become
perpendicular to the inclined ground surface. As discussed by Chuah et al.
(2011), this indicates dominance of vortical effects\ over buoyant uplifting
and links firewhirls with other intensive vortical flows, although the
presence of density variations and buoyancy remains essential in firewhirls.

\begin{figure}
 \centering
\includegraphics[width=7cm]{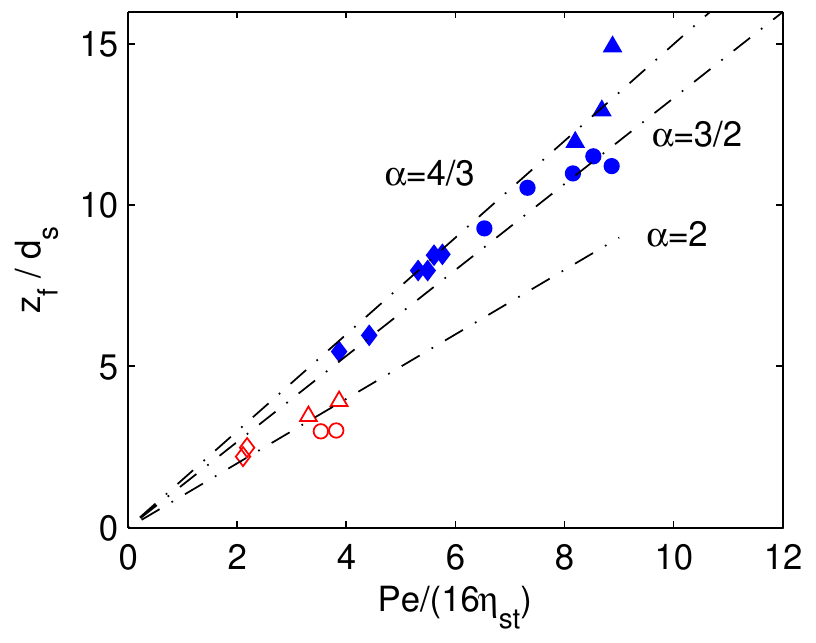} 
\caption{Firewhirl length $z_f$ versus Pe number. Experiments by Chuah et al.
(2011): open symbols -- no rotation; solid symbols
 -- with rotation. Dash-dotted lines -- theory by Klimenko
and Williams (2013) corresponding to $\alpha=4/3$, 3/2 and 2; $d_s$ is
the diameter of the fuel source; $\eta_{st}$ is the stoichiometric value of the mixture fraction.}
\label{fig6}
\end{figure}

Klimenko and Williams (2013) have recently extended the analysis of Kuwana
et al. (2011) and introduced a theory, that uses velocity approximations
based on the compensating regimes and determines the normalised flame length
in terms of the Peclet number and the effective value of the exponent $%
\alpha $. While Klimenko and Williams (2013) take into account the presence
of the viscous/diffusive core, figure \ref{fig6} presents a simplified
treatment linked to the power laws with characteristic values of $\alpha $
used in the rest of the present work: 2, 3/2 and 4/3. The value $\alpha =2$
is associated with flows that have a potential image on the $z$-$r$-plane,
while the compensating range of $4/3\leq \alpha \leq 3/2$ is applicable to
the case when rotation in the flow is strong. While buoyancy prevents vortex
breakdowns and thus favours the exponent of $4/3$ over $3/2,$ the presence
of diffusivity acts in the direction of increasing the effective value of
the exponent $\alpha $. The experimental points of Chuah et al. (2011) shown
in figure \ref{fig6} are in a good agreement with the theoretical
predictions.

\subsection{Supercell tornadoes}

Tornadoes are more affected by atmospheric disturbances than tropical
cyclones and, typically, demonstrate noticeable fluctuations of the axial
vorticity at different elevation levels, while in a strong swirl the
vorticity $\omega _{z}$ is independent of $z$ to the leading order of
approximation. In tornadoes,\ the region of interest\ has a characteristic
AGL (above ground level) of several hundred meters. For the purpose of
comparison, the largest and most stable tornadoes need to be selected as
they are least affected by atmospheric fluctuations, have an axisymmetric
(or near-axisymmetric) structure with uniform distribution of vorticity at
different altitudes and the largest possible range of radii of the
intensification region. As discussed previously, the stages when the axial
vorticity is exhausted in surrounding flow are best described by
conventional Burgers vortices and are different from the intensification
stage considered in the present work. According to Fujita (1981), the
strongest tornadoes reaching F4 grades on the Fujita scale represent less
than 3\% percent of all occurrences of tornadoes while F5 tornadoes are
rare. The strongest and most stable tornadoes with intense rotation are
usually embedded into the core region of a mesocyclone as a part of
supercell thunderstorms.

The exponent $\beta =\alpha -1$ in $v_{\theta }\sim 1/r^{\beta }$ has been
occasionally reported for large tornadoes. Wurman and Gill (2000) presented
high resolution measurements of F4 tornado formed in a supercell storm near
Dimmitt, Texas in 1995 and reported a tangential velocity profile
approximated \emph{within the range of $100\func{m}<r<1$km } by a power law
with $\beta =0.6\pm 0.1.$ Lee and Wurman (2005) presented measurements for a
large F4 tornado that hit a small town of Mulhall in Oklahoma in 1999. The
tornado was unusual: it had a very large core with the radius of maximal
winds $r_{m}$ (RMW) nearly reaching 1km scale \emph{and multiple vortices
orbiting the common core (Wurman, 2002).} The overall slope of axisymmetric
profiles of tangential velocity within the range of 1km to 3km is reported
as having $0.6\leq \beta \leq 0.7,$ although $\beta $ has significant
variations within this range of radii. The slope of the profiles within the
range of $1<r/r_{m}<1.5$ is around $\beta =0.5$ or lower, although increases
in $\beta $ outside the radius of $1.5r_{m}$ (or $2r_{m}$ for some of the
profiles) indicate that axial vorticity was small or negative in this region
-- this corresponds to $\beta \approx 1$. \emph{Wurman (2002) later reported 
$0.5\leq \beta \leq 0.6$ for this tornado.}

\begin{figure}
 \centering
\includegraphics[width=14cm]{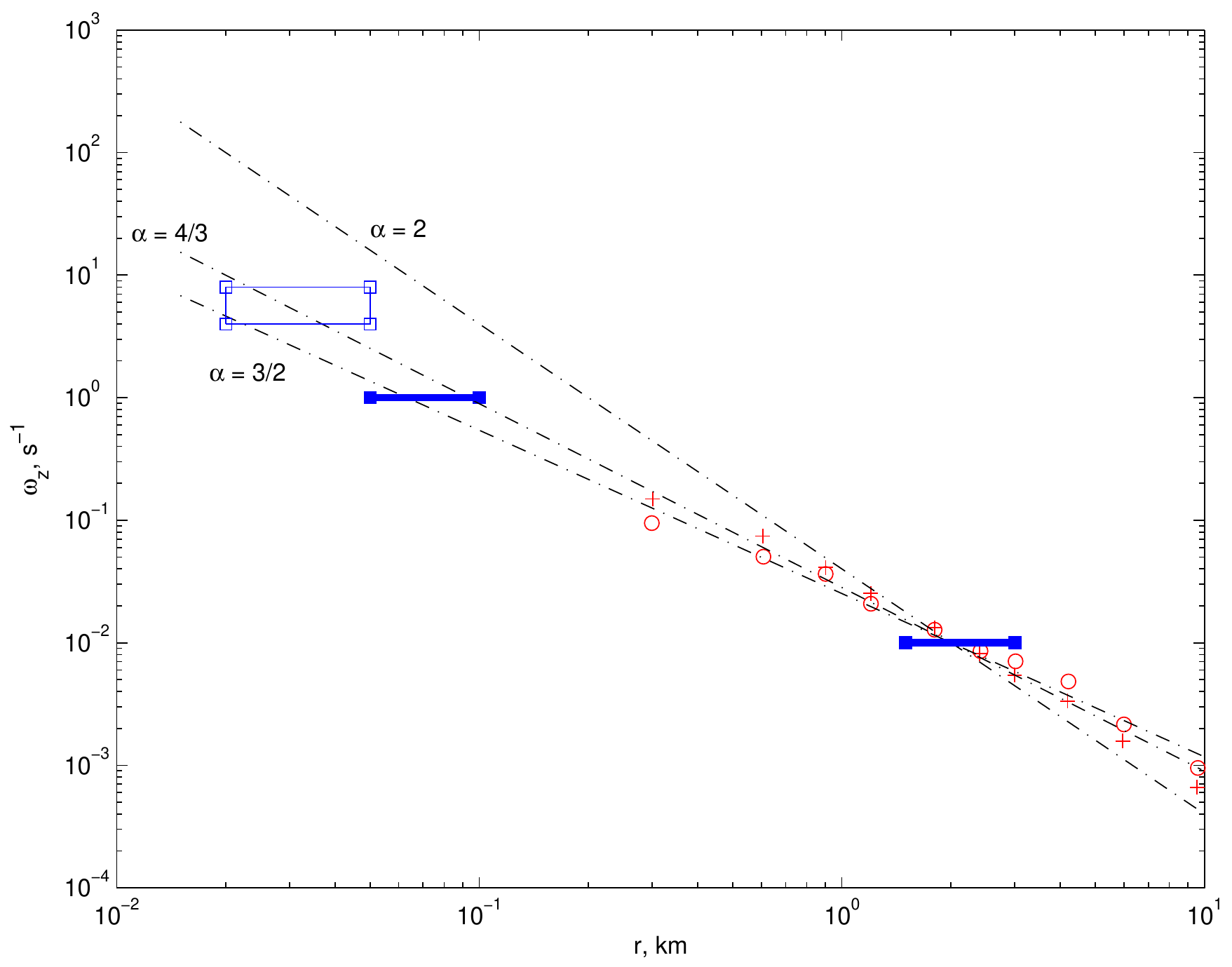} 
\caption{The range of radii characterising the inner and outer
scales of a typical mesocyclone tornado and the corresponding typical levels
of axial vorticity at these scales. The values of typical parameters are
taken from Brooks et al. (1993), Dowell and Bluestein (2002), Bluestein and
Golden (1993) and other publications ($\blacksquare$---$\blacksquare$). Maximal vorticity ever measured in tornadic flows as reported by Wurman (2002) ($\Box$---$\Box$).    Symbols: the scaling of maximal mesocyclonic vorticity versus grid spacing as reported by Cai (2005) for storms at 
Garden City ($+$) and Hays (o).    
The dash-dotted lines demonstrate the slope of $\omega _{z}\sim 1/r^{\alpha }$ for $\alpha =2,$ $3/2,$ $4/3$.}
\label{fig4}
\end{figure}

As mentioned previously, the vorticity-based comparison of theory and
measurements is more direct. The consistency of the power laws of the
compensating regime is now examined in comparison with the conventional
estimations of vorticity levels in tornadic flows. According to reviews of
atmospheric measurements by Brooks et al. (1993), Bluestein and Golden
(1993), and Dowell and Bluestein (2002), a typical supercell tornado
amplifies its axial vorticity from $\sim 0.01$s$^{-1}$ in the outer region
with a span of $3$-$7$km to a level of $\sim 1$s$^{-1}$ in the core of the
tornado with a scale of $\sim 100$m. Figure \ref{fig4} illustrates the
comparison of these parameters with the power laws $\omega _{z}\sim
1/r^{\alpha }.$ The thick lines show the radius of $1.5$-$3.5$km
corresponding to the scale \ $3$-$7$km for the surrounding vorticity of $%
\omega _{z}=0.01$s$^{-1}$ and the radius of $50$-$100$m corresponding to the
scale of $100$-$200$m for the vorticity of $\omega _{z}=1$s$^{-1}$ in the
tornadic core. \emph{\ The box indicates Wurman's (2002) estimate for the
highest vorticity ever measured in tornadoes, which has been detected in
multiple vortices of the Mulhall tornado. This vorticity reached 4--8s$^{-1}$
and within scales of 40--100m.} The compensating exponents $\alpha =4/3$ and 
$\alpha =3/2$ are reasonably consistent with the commonly accepted
characteristics of supercell tornadoes.

\emph{Cai (2005) reported fractal scaling of vortical characteristics of a
mesocyclone, which are expressed as maximal vorticity $\left( \omega
_{z}\right) _{\text{max}}$ measured on a given grid versus grid spacing $%
r_{g}$. The characteristic axial vorticity $\omega _{z}$ at the distance $%
r_{g}$ from the flow convergence centre can serve as an estimate for $\left(
\omega _{z}\right) _{\text{max}}$. Several mesocyclones have been analysed
and it was found that, as expected for fractals, the scaling can be
accurately approximated by the power law $\left( \omega _{z}\right) _{\text{%
max}}\sim 1/r_{g}^{\alpha }$. Cai (2005) found that during intensification
of the Garden City tornadic mesocyclone $\alpha $ increased from 1.31 to
1.59, while for the non-tornadic Hays mesocyclone the value $\alpha $\
ranged only from 1.23 to 1.32 not reaching 4/3 --- the lower boundary of the
compensating exponent predicted by the presented theory. The scaling
profiles for the highest $\alpha $ reported by Cai (2005) are shown in
figure \ref{fig4}. It seems that Cai's fractal method can recover regular
exponents for radii reaching 10km, where the mesocyclonic flow becomes quite
irregular. }

\begin{figure}
 \centering
\includegraphics[width=12cm]{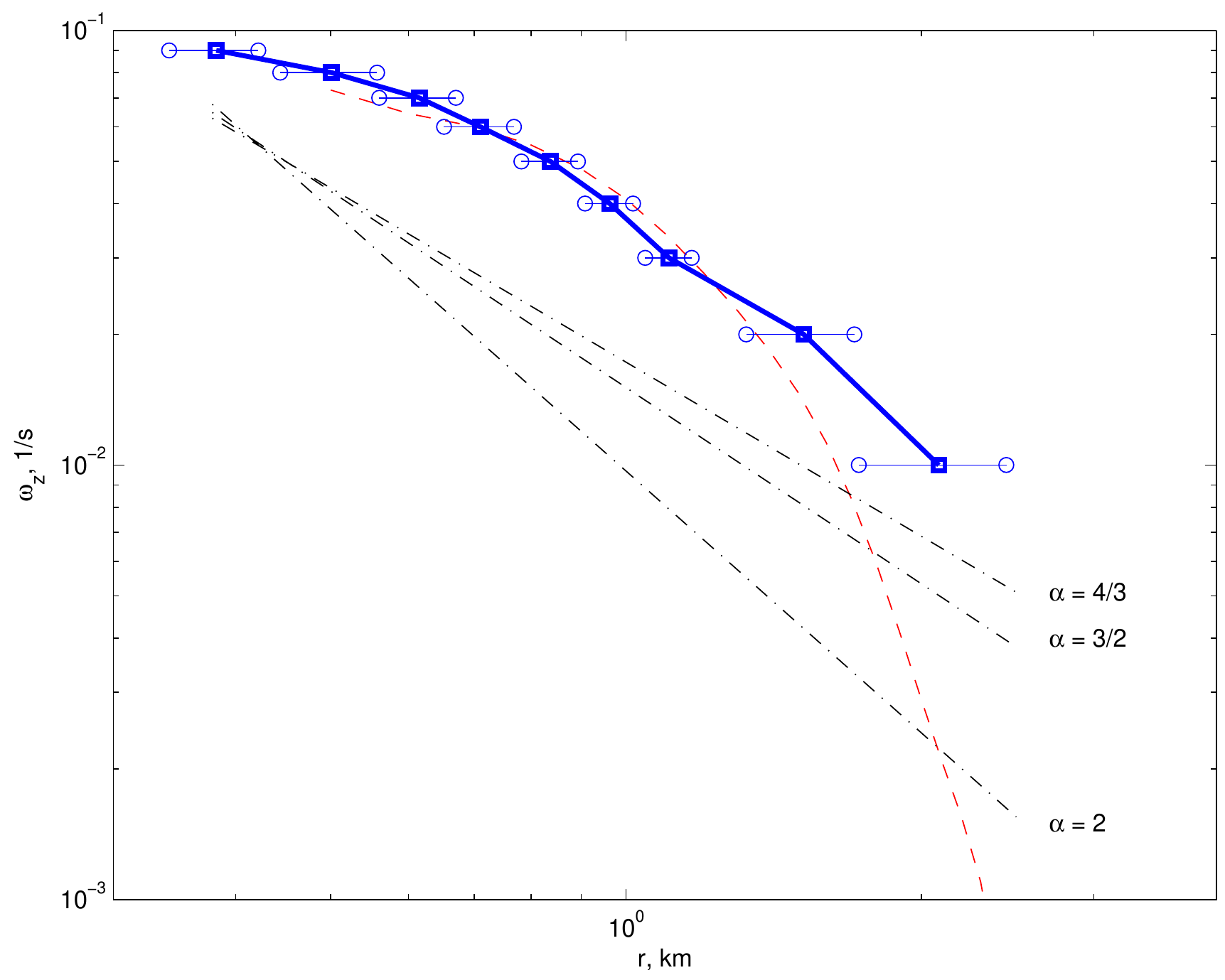} 
\caption{  Axial vorticity $\omega _{z}$ versus radius $r$ for
tornado 4 of the McLean storm at 23:38 UTC. The data are taken from Dowell and
Bluestein (2002): \emph{ the dash line connecting symbols ($\Box$---$\Box$) shows $\omega _{z}$\ obtained
from the axial vorticity contour plots with radius variations shown by the horizontal error bars; the dashed line corresponds to $\omega
_{z}$ evaluated from $\gamma$. The dash-dotted lines display the slopes of $\omega_{z}\sim 1/r^{\alpha }$ for $\alpha =2,$ $3/2,$ $4/3$.} }
\label{fig5}
\end{figure}

Several tornadoes that appeared in the 1995 McLean (Texas) storm were
measured by a Doppler radar (Dowell and Bluestein, 2002) and were also
surveyed and photographed from the ground (Wakimoto et al. 2003). Among
these tornadoes, tornado 4 was the strongest, largest and most stable
tornado, reaching a rating of F4-F5 on the Fujita scale. Unlike the other
tornadoes in this storm, the axial vorticity in tornado 4 was fairly uniform
up to AGL of more than 4 km, it had a regular, nearly axisymmetric shape,
and it persisted for more than an hour. \ The ground damage survey by
Wakimoto et al. (2003) indicates that the radius of maximal winds and the
corresponding damage (F3 at 23:38 UTC, Coordinated Universal Time) did not
exceed $150$m. The axial vorticity has been determined a) from the contour
plot by calculating the average radius of each vorticity contour line and b)
from the reported circulation $\gamma (r)$ under assumptions of an
axisymmetric flow. The results are shown in figure \ref{fig5} for the
tornadic range of radii ($r\leq 4$km) and are reasonably consistent. The
error bars show the standard deviations in evaluating average radii ---
large deviations are indicative of a non-axisymmetric flow. The increasing
difference between the curves at $r>1$km is explained by the difficulty of
evaluating $\omega _{z}(r)$ from $\gamma (r)$ due to an increasingly
non-axisymmetric structure of the flow at mesocylonic scales. Note that the
exponents of the compensating regime $4/3\leq \alpha \leq 3/2$ produce a
reasonable match to the measured vorticity levels.\ 

\emph{\ Dowell and Bluestein (2002) also reported the convergence rates
during formation of tornado 4. According to the analysis of the vortex
evolution in Section 3.5, a constant convergence rate $\lambda $ that
corresponds to initially weak swirl with $\alpha =2$ in (\ref{lam}) is
gradually replaced by the convergence rate increasing towards the axis
according to $\lambda \sim r^{\alpha -2}$ and the value $\alpha $ belonging
to the compensating range. As illustrated in figure \ref{fig2}, this
replacement occurs through extension of the compensating regime towards the
axis. The convergence profiles reported by Dowell and Bluestein (2002) are
consistent with the scheme illustrated in figure \ref{fig2}. }

\subsection{Tropical cyclones}

Tropical cyclones are the largest and most stable vortices observed in the
Earth's atmosphere. The core region of the cyclone consists of the eye
surrounded by the eye wall and, in most cases, has a characteristic radius
of around 20---40km. This region has a noticeably higher temperature and is
strongly affected by buoyancy, while the temperature increments in the
surrounding flow are much smaller. The maximal wind speeds are achieved
within the outer rim of the core. The intensification region, which is
located just outside the core region and above the surface boundary layer,
also involves reasonably strong tangential winds. This region is
characterised by the presence of some updraft flow of air (which is, of
course, weaker than the updraft in the eye wall). The radial winds (and, to
lesser extent, the tangential winds) are affected by the Ekman effect near
the ground or sea surface; hence measurements outside the immediate surface
boundary layer\ should be preferred in the context of our analysis. The
intensification region is limited by its outer radius, which can stretch
beyond 100km. The region located outside the intensification region is also
subject to strong influence from the cyclone. The radius of this region,
which is called here ''peripheral'', can extend to\ 500 km and, possibly,
beyond. The peripheral region can be can be seen as a two-dimensional
vortical sink without any significant updraft. The more remote sections of
this region are affected by fluctuations of synoptic weather patterns.

Approximating the tangential velocity profile in the form of the power law $%
v_{\theta }\sim 1/r^{\beta }$ is conventional in cyclone-related literature.
This power law implicitly assumes that $\gamma _{0}\approx 0$ in (\ref
{alf_bet}) and this may be adequate for many tropical cyclones. Although $%
\beta $ may experience some variations, the estimates $\beta =-1$ for the
core region, $\beta =0.5$ for the intensification region and $\beta =1$ for
the peripheral region are common in the literature (Gray, 1973; Emanuel,
2003). Although $\beta =0.5$ is considered to be the best average
approximation in the intensification region (Gray, 1973; Emanuel, 2003),
some estimates of $\beta $ can deviate from this value. For example, one of
the early works by Hughes (1952) nominated $\beta =0.62$ as the best fit to
data obtained from a number of reconnaissance flights into cyclones (these
flights began in 1943 and represent the most important source of information
about hurricanes) and he noted that this exponent is reasonably close to the
more conventional value of $\beta =0.5$. Riehl (1963) observed evolution of
unusual tangential wind profiles in hurricanes Carrie (1957) and Cleo
(1958)\ relaxing towards profiles with $\beta \approx 0.5$.

Explaining the value of $\beta =0.5$ in the intensification region is not
trivial. Riehl (1963) demonstrated that $\beta =0.5$ produces a good fit for 
$v_{\theta }\sim 1/r^{\beta }$ in six different hurricanes. He noted that
assuming both the moment of the tangential component of the surface stress $%
r\sigma _{\theta }$ and the drag coefficient $C_{D}$ to be independent of $r$
is sufficient (but not necessary) for $\beta $ to be $0.5.$\ Although Pearce
(1993) put forward arguments supporting this assumption, the independence of 
$C_{D}$ from $r$ is generally not supported by the measurements. The data
reported by Hawkins and Rubsam (1968) and by Palmen and Riehl (1957)
indicate, however, that $C_{D}\sim 1/r^{\zeta }$\ with $\zeta $ ranging
between 0.4 and 0.7 while Palmen and Riehl (1957) determined that, on
average, $r\sigma _{\theta }\sim 1/r^{0.6}$. The approach of the present
work indicates that, while losses of angular momentum are important in
intensive vortices, the flow adjusts itself to compensate for disturbances
and relax towards the exponents of the compensating regime.\ In his
thermodynamic theory of steady tropical cyclones, Emanuel (1986)
demonstrated that $\beta \approx 0.5$ just outside RMW is consistent with
typical temperatures on the sea surface and in tropopause. Here, one can
note that $\alpha =\beta +1=1.5$ is the same as the value $\alpha ^{\ast
}=3/2$ suggested in Section 3 for the compensating regime.

Hawkins \& Rubsam (1968) and Hawkins \& Imbembo (1973) reported axial
vorticity distributions and other characteristics for two hurricanes, Hilda
(1964) and Inez (1966). These distributions do not show any significant
dependence on $z$ at lower altitudes, although the axial vorticity profile
of Hilda had an irregularity at the altitudes above two kilometers, while $%
\omega _{z}$ in Inez remained regular up to the altitudes of four
kilometers. Vorticity distributions in these and other hurricanes tend to be
reasonably consistent with the strong swirl approximation and the exponents
of the compensating regime.

\begin{figure}
 \centering
\includegraphics[width=14cm]{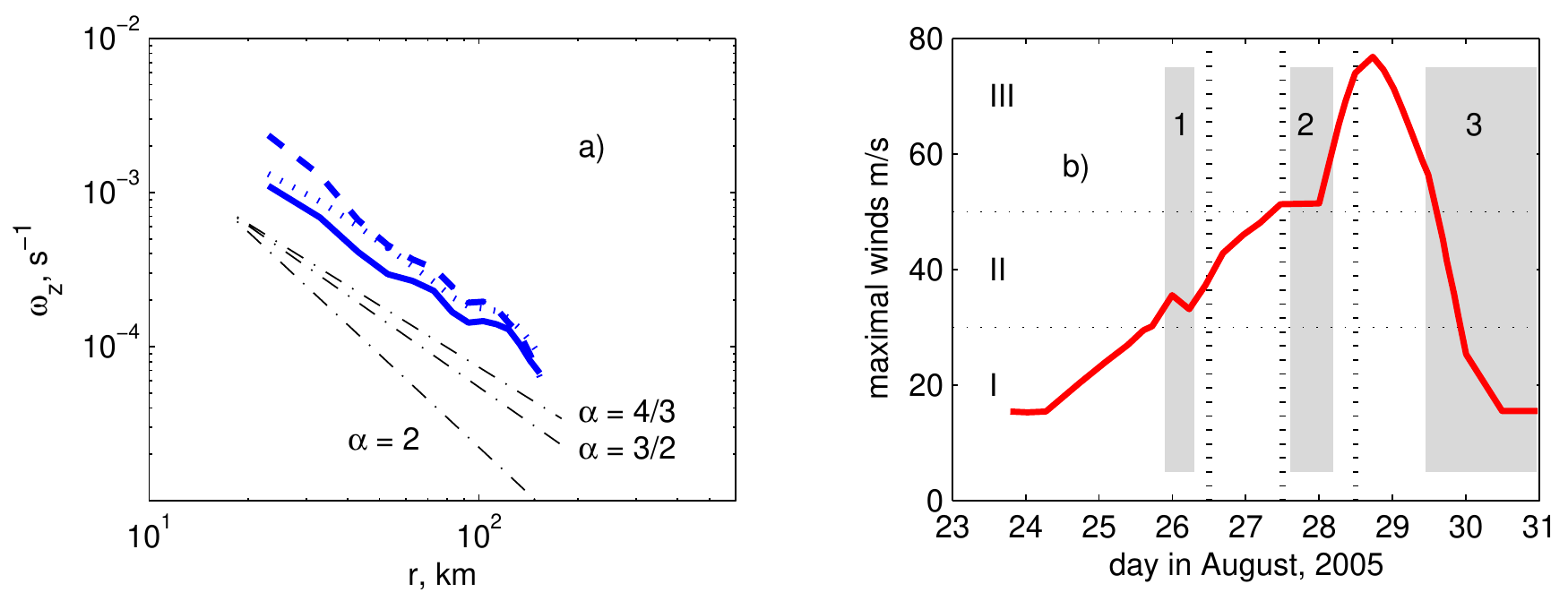} 
\caption{ \emph{Hurricane Katrina, 2005. 
a) Axial vorticity $\omega _{z}$ versus radius $r$ measured 
at 12:00 26 August  UTC (------); 
at 12:00 27 August  UTC (......); 
at 12:00 28 August  UTC (-- -- -- --). 
The thin dash-dotted lines show the slope of $\omega _{z}\sim 1/r^{\alpha }$ for $\alpha =2,$ $3/2,$ $4/3$. 
b) Maximal winds versus versus UTC dates. 
The ranges for I -- prehurricanes, II -- minimal hurricanes and III -- major hurricanes are shown (see table 1 for the definitions). 
The vertical dashed lines correspond to the three time moments listed above. 
The gray areas indicate  events disturbing the state of the hurricane: 
1 -- the first landfall over southern Florida, 2 -- eye replacement cycle, 3 -- the second landfall.
The data and information are taken from Knabb et al. (2005) and Powell et al. (2010). The axial vorticity is evaluated from the data set 
provided by Hurricane Research Division of the US National Oceanic and Atmospheric Administration (HRD-AOML-NOAA). }}    	
\label{fig7}
\end{figure}

\begin{figure}
 \centering
\includegraphics[width=14cm]{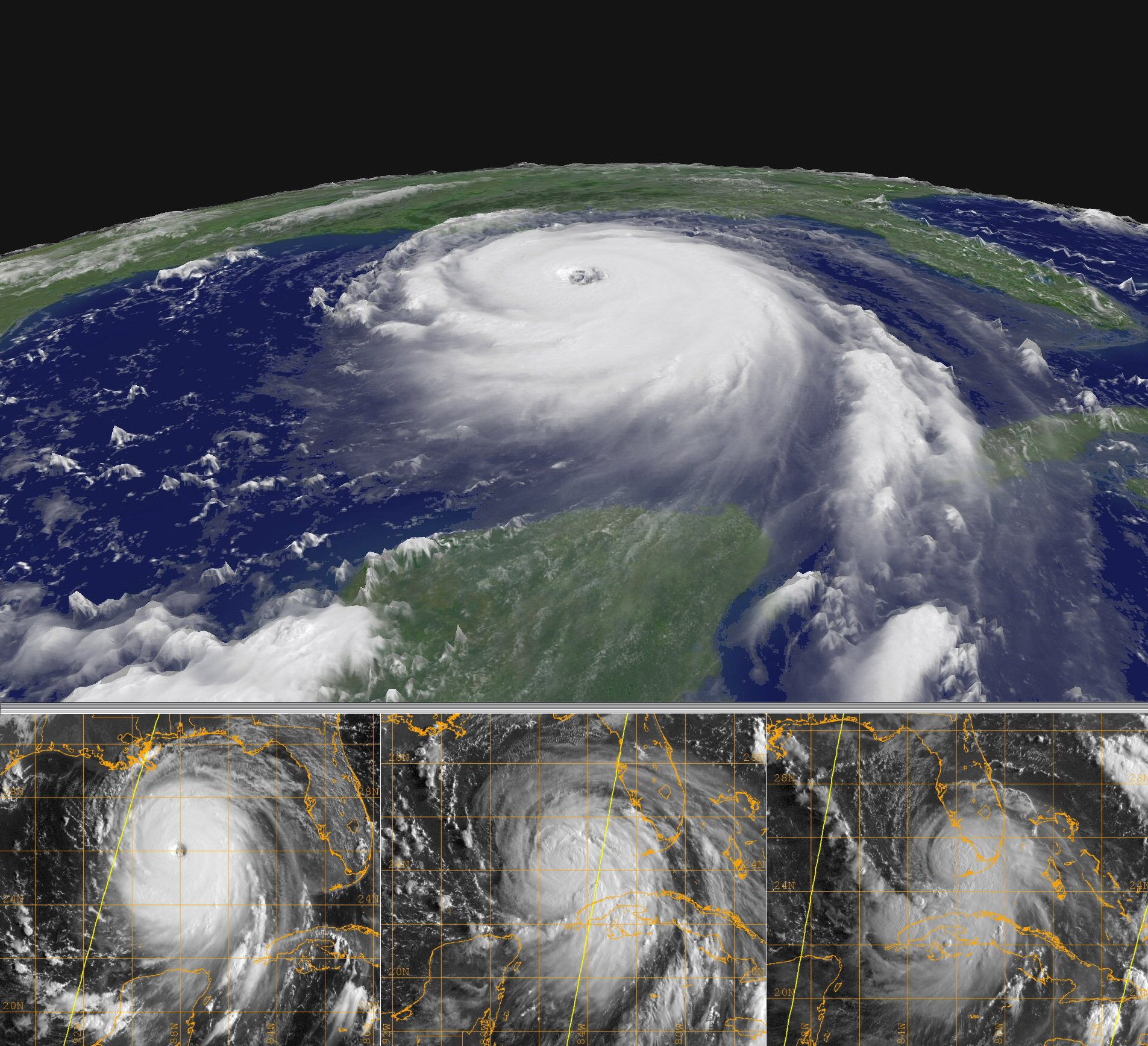} 
\caption{ \emph{Satellite photographs of hurricane Katrina. Top: taken at 15:45 UTC on 28 August, 2005, when the hurricane was about to reach its maximal strength (courtesy of the NASA Goddard Space Flight Center). Bottom, from right to left: images taken on 26, 27 and 28 August at approximately   the same times (i.e. 12:00 UTC) as for the data presented in figure \ref{fig7}a (courtesy of the US Naval Research Laboratory) } }
\label{fig8}
\end{figure}

\emph{Katrina (2005) is one of the strongest hurricanes on record to hit the
American continent. After passing over the southern tip of Florida (zone 1
in figure \ref{fig7}b), hurricane Katrina quickly regained its strength. On
27 August, Katrina reached the warm waters of the Loop Current of Mexican
Gulf\ (Powell et al., 2010) and became a major hurricane (see table 1) but
its further strengthening, was delayed due to adjustments that are likely to
have been caused\ by the eyewall replacement cycle (Knabb et al., 2005,
Houze et al., 2007). After completion of the cycle (shown as zone 2 in
figure \ref{fig7}b), Katrina intensified at an extremely high rate, reaching
category 5 on 28 August. The inclined satellite photo in figure \ref{fig8}
(top), taken when the hurricane was approaching its maximal strength, shows
the large scale of the hurricane and a very distinct eye that forms a
depression reaching a diameter exceeding 50km. This clearly visible eye is
indicative of axial downdrafts and is typical of major hurricanes, although
it seems that the formation of a prominent eye in Katrina was delayed by the
eyewall replacement cycle until the early morning of 28 August (although the
eye can be detected in infrared satellite images taken on 27 August --- see
Knabb et al., 2005).\ Katrina started to reduce its strength towards the end
of 28 August, and on 29 August it made its second landfall on the Louisiana
coast causing flood and devastation (shown as zone 3 in figure \ref{fig7}b).
In the following days, Katrina quickly lost its might but, as a tropical
depression, reached as far as the states near the Great Lakes and caused
rains in Canada. The axial vorticity profiles are evaluated from the
hurricane wind speeds and shown at 12:00 UTC on 26, 27 and 28 August, when
Katrina was only a minimal hurricane, just reached the status of a major
hurricane and was a major category 5 hurricane close to its peak state. The
three bottom images in figure\ \ref{fig8} illustrate the state of the
hurricane at the time of the measurements. Only last of the curves (i.e.
measured on 28 August) presented in figure \ref{fig7}a corresponds to the
vortex with a large visible eye present. The profiles are generally
consistent with the compensating exponents. The slope of the vorticity
curves tends to increase when the hurricane becomes stronger. }

Mallen et al. (2005) presented a comprehensive analysis of axisymmetric
tangential velocity and axial vorticity distribution in tropical storms
involving 251 different cases. The results are summarised in table 1. The
scaling exponents were determined in the region between $1\leq r/r_{m}\leq $ 
$3$ where $r_{m}$ denotes RMW. The exponents reported for different storms
indicate a significant scattering with $\alpha $ ranging from 1.05 to 1.7.
The best approximation for the exponent ($\alpha =1.37$) was determined as
the average over all storms. Mallen et al. (2005) also found that the value
of the exponent correlates with the strength of the storms and divided all
storms into three classes: prehurricanes, minimal hurricanes and major
hurricanes. The average value of $\alpha $ for each of the classes were
determined to be 1.31, 1.35 and 1.48. The higher values of $\alpha $
correspond to stronger storms. These profiles also have differences at $%
r=r_{m}$ where the slope of these profiles is much steeper than that
predicted by $\omega _{z}\sim 1/r^{\alpha }$ due to the dominance of
vorticity $\gamma _{0}$ accumulated within the core. As expected from the
present analysis, the average vorticity profiles reported by Mallen et al.
(2005) for pre-hurricanes and minimal hurricanes are flatter at $r>r_{m}$
but are nevertheless steeper at $r=r_{m}$ indicating a stronger influence of 
$\gamma _{0}$ on the flow just outside RMW. The major hurricanes, which
belong to category three and above, usually have a clearly visible eye with
a cloud clearance created by downdrafts. As discussed in Section 3.5, this
corresponds to reduced influence of the core and to a compensating exponent
of 3/2. Note that the range of $\alpha =$ 1.31, 1.35 and 1.48 is very close
to the range of $4/3\leq \alpha \leq 3/2$ predicted by the present analysis
of the compensating regime.


\begin{table}
\caption{ \emph{Comparison of average values of the exponent $\protect\alpha$
determined by Mallen et. al. (2005) for different hurricane classes with
theoretical predictions; $n$ is the number of cases analysed; $ \Delta\alpha = \alpha - \left< \alpha \right> $ is deviation of $\alpha$ from its average;  $v_m = v_{\theta }\left( r_m \right) $ is the speed of maximal winds; the averages in the last line are weighted by $n$. 
(Note that the hurricanes of category 1 on the Saffir-–Simpson scale have the maximal winds of at least 33m/s, which is very close to the 30m/s threshold for the minimal hurricanes.) 
}}
\bigskip 
\begin{center}
\begin{tabular}{p{3cm}cccccc}
\toprule
& $ v_m $ & \bf{category} & $n$ & \multicolumn{2}{c}{ \bf{measurements}} & \bf{theory}  \\ 
 \cmidrule(r){5-6}   \cmidrule(r){7-7}
&  &  &  &  $ \left< \Delta\alpha^{2} \right>^{\frac{1}{2}} $ &  $ \left< \alpha \right> $ &  $\alpha^{*} $ \\ 
\midrule

\bf{pre-hurricane}  & \TEXTsymbol{<}30m/s & \tiny{tropical storm} & 73 &  0.12 & 1.31  & $ 4/3\approx 1.33$ \\  [-1ex]
 & & \tiny{or depression} &  &  &  &  \\  [1ex]
\bf{minimal hurricane} & 30-50m/s & 1, 2 & 106 & 0.14 & 1.35  & $4/3\approx 1.33$ \\ [2ex] 
\bf{major hurrricane} & \TEXTsymbol{>}50m/s & 3, 4, 5 & 72 &  0.11 & 1.48  & $3/2 = 1.50$\\ 
\midrule
\bf{average} & \bf{all} & \bf{all} & \bf{251} & \bf{0.14} & \bf{1.37} & \bf{1.38} \\
\bottomrule
\end{tabular}
\end{center}
\bigskip
\end{table}

\bigskip

\section{Conclusions\protect\nopagebreak 
}

\emph{The present work develops a theory of intensive vortices that are
distinguished by a fluid flow from peripheral to central regions and a
significant amplification of rotational motion near the centre of the flow.
The theory is generic and based on the strong swirl asymptotic appoximation,
considered from the perspective of vorticity equations. Hurricanes,
tornadoes and firewhirls, which are also examined in the present work, are
well-known examples of intensive vortices. Conventional axisymmetric
vortical schemes that imply a potential flow image on the axial-radial plane
(such as the Burgers vortex) do not represent a good model for an intensive
vortex with significant ambient vorticity and strong swirl. In terms of the
power law $\psi \sim r^{\alpha }z$, flows with a potential $r$-$z$-image
correspond to $\alpha =2,$ while the present theory of intensive vortices
suggests that the exponent $\alpha $ should reach its compensating values $%
\alpha ^{\ast }$ lying in the range of $4/3\leq \alpha ^{\ast }\leq 3/2$.
This exponent is expected to be valid outside the core extending outward to
the intensification region, where updrafts amplifying the axial vorticity
are significant. The compensating values of the exponent are determined by
consistency of velocity/vorticity interactions that, in axisymmetric
conditions considered here are, controlled by the vortical swirl ratio $K$.
This parameter $K=(S/\func{Ro})^{1/2}$ represents the geometric mean of two
conventional parameters --- the swirl ratio and the inverse Rossby number. }

\emph{While intensive vortices tend to evolve slowly, they are still
inherently non-stationary and evolutionary aspects of these vortices need to
be considered. Formation of the vortex involves appearance of strong swirl
condition at a distance from the centre followed by centripetal propagation
of these conditions. In the regions where the swirl becomes strong, the
exponent $\alpha $ relaxes to its compensating range. This scheme is
different from the centrifugal propagation of these conditions considered by
Klimenko (2001a). Two aspects of the influence of viscosity on the core of
the vortex are of interest. First, the value of $K$ in the viscous core is
higher than in the surrounding flow, which creates conditions for the vortex
breakdown in the core. Second, viscosity is shown to remove the singularity
of the compensating exponents near the axis. }

\emph{Interactions of velocity and vorticity are generally known to have a
destabilising effect in most of the fluid flows. In intensive vortices,
however, these interactions enact a stabilising mechanism that compensates
for possible variations of the vortical swirl ratio and, as the fluid flows
towards the axis, relaxes the exponents to their compensating range $4/3\leq
\alpha ^{\ast }\leq 3/2$. Existence of this stabilising mechanism explains
the persistent character of the intensive vortices.\ The compensating
exponents can be seen as equilibrium values --- the actual exponents
measured in the specific vortices may deviate from but tend to relax to
these equilibrium values. In the atmosphere, intensive vortices are
continuously disturbed by changes in surrounding atmospheric and surface
boundary conditions. The measurements presented here indicate a reasonable
but not absolute agreement with the theory when specific cases are analysed.
However, when the averages are evaluated over a large set of experiments
(251 hurricanes analysed by Mallen et al. 2005) and the disturbances and
variations of conditions are effectively removed, the match between
theoretical predictions and experiments becomes very accurate. }

\section*{Acknowledgments}

\emph{The author thanks Stewart Turner for discussion and advice. The author
appreciates assistance and information he has received from NASA. The data
set for hurricane Katrina is provided by the Hurricane Research Division of
the US National Oceanic and Atmospheric Administration. The satellite images
are the courtesy of NASA and the US Naval Research Laboratory. The author's
work is supported by the Australian Research Council. }

\appendix

\section{Viscous core in the strong swirl approximation}

\emph{This section proves that the singularity of the compensating regime is
removed by viscosity near the axis, and finds the corresponding consistent
asymptotic at }$r\rightarrow 0.$ In the viscous core, the influence of
viscosity is significant and the characteristic radius $r_{\ast }=\nu
/(Lv_{\ast })$ is determined by $\func{Re}=1$. \ Since $\func{Re}_{z}\equiv
z_{\ast }v_{\ast }/\nu =\func{Re}/L\gg 1$, it is assumed here that $L\sim 1/%
\func{Re}_{z}\ll 1.$ Only the leading terms with respect to $L$ need to be
considered here. Equations (\ref{AA0def})-(\ref{AA0G1}) can be simplified 
\begin{equation}
\Gamma _{00}=\Gamma _{00}(R,T),\ \Gamma _{01}=\Gamma _{01}(R,T),\;\;\Omega
_{r00}=\Omega _{r01}=0,\ \ \Gamma _{10}=0,
\end{equation}
\begin{equation}
\Psi _{00}=F_{0}\left( R,T\right) +F_{1}(R,T)Z,\ V_{r10}=V_{r01}=0,\ \Psi
_{10}=\Psi _{01}=0,  \label{apa_F}
\end{equation}
\begin{equation}
V_{r00}\frac{\partial \Gamma _{00}}{\partial R}=R\frac{\partial }{\partial R}%
\left( \frac{1}{R}\frac{\partial \Gamma _{00}}{\partial R}\right) ,\ \ \frac{%
\partial \Gamma _{00}}{\partial T}+V_{r00}\frac{\partial \Gamma _{01}}{%
\partial R}=R\frac{\partial }{\partial R}\left( \frac{1}{R}\frac{\partial
\Gamma _{01}}{\partial R}\right) ,
\end{equation}
\begin{equation}
2\frac{\Gamma _{00}}{R^{3}}\Omega _{r11}=-\frac{D_{00}\Omega _{\theta 00}/R}{%
DT}+\frac{1}{R}\frac{\partial }{\partial R}\left( \frac{1}{R}\frac{\partial
\Omega _{\theta 00}R}{\partial R}\right) ,  \label{apaE3}
\end{equation}
\begin{equation}
V_{r11}\frac{\partial \Gamma _{00}}{\partial R}=-\frac{D_{00}\Gamma _{11}}{DT%
}+R\frac{\partial }{\partial R}\left( \frac{1}{R}\frac{\partial \Gamma _{11}%
}{\partial R}\right) ,  \label{apa_v1}
\end{equation}
These equations can be integrated, resulting in 
\begin{equation*}
V_{r00}\Omega _{z00}=\frac{\partial \Omega _{z00}}{\partial R},\ \;\Omega
_{z00}=\Omega _{z}^{\circ }\exp \left( \int_{0}^{R}V_{r00}\mathrm{d}R\right)
,\ \ \Gamma _{0i}=\int_{0}^{R}\Omega _{z0i}R\mathrm{d}R,\ 
\end{equation*}
\begin{equation}
\frac{1}{R}\frac{\partial \Gamma _{00}}{\partial T}+V_{r00}\Omega _{z01}=%
\frac{\partial \Omega _{z01}}{\partial R},\text{ \ }\Omega _{z01}=\Omega
_{z00}\int_{0}^{R}\frac{\partial \Gamma _{00}}{\partial T}\frac{\mathrm{d}R}{%
\Omega _{z00}R},  \label{E3_9}
\end{equation}
where $\ i=0,1$. Note that in the viscous case the value of the vortical
swirl ratio in the core denoted here by $\tilde{K}$ increases 
\begin{equation}
\tilde{K}^{2}=K_{r}^{2}(r_{\ast })=\left( \frac{\gamma \omega _{z}}{v_{z}^{2}%
}\right) _{r=r_{\ast }}\sim \frac{K^{2}}{\func{St}},  \label{E_par}
\end{equation}
since $\Omega _{z00}\neq 0$ there (unlike in the inviscid case where $\Gamma
_{00}=\Gamma _{00}(T)$ and $\Omega _{z00}=0)$. Here, $K$ and $\func{St}$
refer to the corresponding values of parameters introduced for the inviscid
flow.

The complete solution $\Psi (R,Z)$ within the core depends on specific
boundary conditions imposed on the flow at large $Z$ and, generally, cannot
be determined without specifying these conditions\ (Turner, 1966). At the
same time, the near-axis behaviour of the stream function is constrained by
a number of consistency conditions and, as demonstrated below, can be
determined by a generic asymptotic analysis involving higher-order terms.
Since $Z=0$ represents a streamline in bathtub-type flows, it is concluded
that $F_{0}=0$ in (\ref{apa_F}). The exponent $\alpha _{0}$ in the asymptote 
$F_{1}\rightarrow C_{0}R^{\alpha _{0}}$ as $R\rightarrow 0$ remains unknown
a priori. The stream function, velocities and circumferential vorticity are
then given by 
\begin{equation*}
\Psi _{00}\rightarrow C_{0}R^{\alpha _{0}}Z,\;\ \ V_{r00}\rightarrow
-C_{0}R^{\alpha _{0}-1},
\end{equation*}
\begin{equation}
V_{z00}\rightarrow \alpha _{0}C_{0}R^{\alpha _{0}-2}Z,\;\ \ \Omega _{\theta
00}\rightarrow -\alpha _{0}(\alpha _{0}-2)C_{0}R^{\alpha _{0}-3}Z\;.
\label{E3_11}
\end{equation}
The value of $\Omega _{r11}$ is determined from (\ref{apaE3}) and then
integrated over $Z$ and multiplied by $R$\ to obtain $\Gamma _{11}$
according to (\ref{AA0def}) 
\begin{equation}
\Gamma _{11}\rightarrow -\frac{C_{0}Z^{2}}{2\Omega _{z}^{\circ }}\alpha
_{0}(\alpha _{0}-2)\left[ \underset{\text{convection}}{\underbrace{%
4C_{0}R^{2\alpha _{0}-4}}}-\underset{\text{viscosity}}{\underbrace{(\alpha
_{0}-2)(\alpha _{0}-4)R^{\alpha _{0}-4}}}\right] \;.  \label{E_gam2}
\end{equation}
The term $V_{r11}$ is determined from (\ref{apa_v1}) then integrated over $Z$
and multiplied by $R$\ to obtain $\Psi _{11}$ according to (\ref{AA0def}) 
\begin{equation}
\Psi _{11}\rightarrow \frac{-C_{0}Z^{3}}{6(\Omega _{z}^{\circ })^{2}}\left[
\alpha _{0}(\alpha _{0}-2)^{2}(\alpha _{0}-4)^{2}(\alpha _{0}-6)R^{\alpha
_{0}-6}+...\right] \;.  \label{E_Fi1}
\end{equation}
Only $\alpha _{0}=2$ can comply with (\ref{E_gam2})-(\ref{E_Fi1}) and other
physical requirements. Indeed, any value above $\alpha _{0}=2$ results in $%
V_{z00}\rightarrow 0$ at the axis and this is not what can be expected in a
bathtub-type flow. Any value $\alpha _{0}<2$ (but not $\alpha _{0}=0$)
results in $\Gamma _{11}\rightarrow \infty $ as $R\rightarrow 0,$ which is
inconsistent with the asymptotic expansion for $\Gamma $ in (\ref{A0ser})
and (\ref{AAser}). Physically, this means that the vorticity $\Omega
_{\theta 00}$ generated by the flow when the circulation is restricted at
the axis is not sufficient to sustain the singularity of $\alpha <2$. The
value of $\alpha _{0}=0$ is also not suitable for this flow since it
requires a mass sink at the axis and $V_{r00}\rightarrow \infty $ as $%
R\rightarrow 0$. Thus, it follows that $\alpha _{0}=2$ in the inner sublayer
of the viscous core.

Regularity of the solution at the axis is now proved but, since equation (%
\ref{E_gam2}) is nullified by $\alpha _{0}=2$, the higher-order terms in the
expansion $\Psi _{00}=\Sigma _{i}C_{i}R^{\alpha _{i}}$ have to be considered
to obtain the asymptotic behaviour of $\Gamma _{11}$\ at the axis. One can
note that $\alpha _{i}$ distinct from $2$, $4,$ $6$ and\ less than $8$
generates $\Gamma _{11}$ and $\Psi _{11}$ exceeding $\Gamma _{00}\sim R^{2}$
and $\Psi _{00}\sim R^{2}$ as $R\rightarrow 0$. Thus, the stream function $%
\Psi _{00}$ is sought in the form of the expansion 
\begin{equation}
\Psi _{00}=Z\left[ C_{0}R^{2}+C_{1}R^{4}+C_{2}R^{6}+C_{3}R^{8}\right]
+O(R^{10})  \label{E5_FI0}
\end{equation}
--- all these terms are actually needed for correct evaluation of the
asymptotes of $\Gamma _{11},$ $\Psi _{11}$ and $V_{z11}$ at the axis.
Equations (\ref{E3_11}) can be used for evaluation of $V_{z00}$, $V_{r00}$
and $\Omega _{\theta 00}$ term by term due to linear character of the
operators in (\ref{AA0def}). The expansions for $\Omega _{z00}$ and $\Gamma
_{00}$ are obtained from (\ref{E3_9}) then substituted into (\ref{apaE3})
and this determines $\Omega _{r11}$, $\Gamma _{11}$, $V_{r11}$ and $\Psi
_{11}$ by (\ref{AA0def}) and (\ref{apa_v1}) 
\begin{equation}
\Omega _{z00}=\Omega _{z}^{\circ }\left( 1-\frac{C_{0}}{2}R^{2}\right)
+O(R^{4}),\ \ \ \Gamma _{00}=\Omega _{z}^{\circ }\left( \frac{R^{2}}{2}-%
\frac{C_{0}R^{4}}{8}\right) +O(R^{6}),
\end{equation}
\begin{equation}
\Gamma _{11}=2\frac{Z^{2}}{\Omega _{z}^{\circ }}R^{2}\left[
48C_{2}-4C_{0}C_{1}+\left(
288C_{3}-8C_{1}^{2}-C_{0}^{2}C_{1}+12C_{0}C_{2}\right) R^{2}\right]
+O(R^{6}),
\end{equation}
\begin{equation}
\Psi _{11}=128\frac{Z^{3}}{(\Omega _{z}^{\circ })^{2}}\left( \frac{C_{1}^{2}%
}{3}-12C_{3}\right) R^{2}+O(R^{4})\;.  \label{E5_FI}
\end{equation}
Note that the corresponding axial velocity 
\begin{equation}
V_{z11}=256\frac{Z^{3}}{(\Omega _{z}^{\circ })^{2}}\left( \frac{C_{1}^{2}}{3}%
-12C_{3}\right) +O(R^{2})  \label{E5_Vz}
\end{equation}
can become negative when $C_{3}$ is sufficiently large.

\section{Vorticity evolution in inviscid axisymmetric flow}

Consider unsteady convection of the initially uniform axial vorticity $%
\omega _{z}=\omega _{0}=\func{const}$ by an inviscid flow with the stream
function given by $\psi =c_{0}r^{\alpha }z$ as in (\ref{21psi}). The
Lagrangian trajectories $r_{t}=r_{t}(t)$ and $z_{t}=z_{t}(t)$ with initial
conditions $r_{t}(t_{0})=r_{0}$ and $z_{t}(t_{0})=z_{0}$ are evaluated by
integration of $\mathrm{d}r_{t}/\mathrm{d}t=v_{r}$ and $\mathrm{d}z_{t}/%
\mathrm{d}r_{t}=v_{z}/v_{r}$: 
\begin{equation}
\frac{\omega _{zt}}{\omega _{0}}=\frac{z_{t}}{z_{0}}=\left( \frac{r_{t}}{%
r_{0}}\right) ^{-\alpha },\;\;\phi (r_{0})-\phi (r_{t})=\tau ,
\end{equation}
where 
\begin{equation}
\tau \equiv c_{0}(t-t_{0}),\ \ \phi (r)\equiv \left\{ 
\begin{array}{cc}
\ln (r), & \alpha =2 \\ 
r^{2-\alpha }/(2-\alpha ), & 0\leq \alpha <2
\end{array}
\right\} \;.
\end{equation}
In evaluation of the Lagrangian value of axial vorticity $\omega
_{zt}=\omega _{zt}(t)$ from the initial condition $\omega
_{zt}(t_{0})=\omega _{0},$ the fact that the vortical lines are frozen into
inviscid flows is used. Substitution of the ratio $r_{t}/r_{0}$ evaluated
from the second equation results in 
\begin{equation}
\frac{\omega _{z}}{\omega _{0}}=\left\{ 
\begin{array}{cc}
\exp (2\tau ), & \alpha =2 \\ 
\left( 1+(2-\alpha )\tau r^{\alpha -2}\right) ^{\alpha /(2-\alpha )}, & 
0\leq \alpha <2
\end{array}
\right\} ,  \label{apb_ww}
\end{equation}
\begin{equation}
\gamma =\left\{ 
\begin{array}{cc}
\frac{\omega _{0}}{2}r^{2}\exp (2\tau ), & \alpha =2 \\ 
\frac{\omega _{0}}{2}\left( (2-\alpha )\tau +r^{2-\alpha }\right)
^{2/(2-\alpha )}, & 0\leq \alpha <2
\end{array}
\right\} \;.
\end{equation}
There is an essential difference between these equations: the second
equation of (\ref{apb_ww}) does approach the quasi-steady solution $\omega
_{z}\sim r^{-\alpha }$ for sufficiently large $t-t_{0}$ or sufficiently
small $r$ while, in\ the first equation of (\ref{apb_ww}), the vorticity
remains $\omega _{z}=\omega _{z}(t)$\ and\ does not become quasi-steady at
any time. For the case of $0\leq \alpha <2$, the quasi-steady (long-term)
asymptotic for $\omega _{z}$ is given by 
\begin{equation}
\frac{\omega _{z}}{\omega _{0}}=((2-\alpha )\tau )^{\alpha /(2-\alpha
)}r^{-\alpha }+...,\ \ \ 0\leq \alpha <2\;.
\end{equation}
It is worthwhile to note that the quasi-steady asymptotes for $\omega _{z}$
are determined by the continuing vertical stretch of the vortex lines and do
not depend on the initial conditions (provided $0<\alpha <2$). The equations
introduced here can be generalised for $c_{0}=c_{0}(t)$ by redefining $\tau $
as 
\begin{equation}
\tau =\int_{t_{0}}^{t}c_{0}(t)\mathrm{d}t\;.
\end{equation}


\newpage

\newpage

\end{document}